\documentclass[aps,prb,twocolumn,superscriptaddress,showpacs]{revtex4-1}
\usepackage{graphicx}
\usepackage{mathrsfs}
\usepackage{bm}
\usepackage{amsfonts}
\usepackage{amsmath}
\usepackage{color}
\usepackage{dcolumn}
\usepackage{epstopdf}
\usepackage{dsfont}
\usepackage{amssymb}
\usepackage{tabularx}
\usepackage{array}
\usepackage{float}
\usepackage{colordvi}
\usepackage{verbatim}
\usepackage{hyperref}
\hypersetup{
	colorlinks=true,
	linkcolor=blue,
	filecolor=blue,
	urlcolor=blue,
	citecolor=blue,
}
\usepackage{amsfonts}
\usepackage{multirow}
\usepackage{bigstrut}
\usepackage{rotating}
\usepackage{booktabs}
\usepackage{braket}
\usepackage{enumerate}

\begin{document}
\title{Engineering second-order topological insulators via coupling two first-order topological insulators}

\author{Lizhou Liu$^\ddag$}
\affiliation{College of Physics, Hebei Normal University, Shijiazhuang, Hebei 050024, China}
\author{Jiaqi An$^\ddag$}
\affiliation{International Centre for Quantum Design of Functional Materials, CAS Key Laboratory of Strongly-Coupled Quantum Matter Physics, and Department of Physics, University of Science and Technology of China, Hefei, Anhui 230026, China}
\author{Yafei Ren}
\affiliation{Department of Physics, University of Delaware, DE 19716, USA }
\author{Ying-Tao Zhang}
\email[Correspondence author:~~]{zhangyt@mail.hebtu.edu.cn}
\affiliation{College of Physics, Hebei Normal University, Shijiazhuang, Hebei 050024, China}
\author{Zhenhua Qiao}
\email[Correspondence author:~~]{qiao@ustc.edu.cn}
\affiliation{International Centre for Quantum Design of Functional Materials, CAS Key Laboratory of Strongly-Coupled Quantum Matter Physics, and Department of Physics, University of Science and Technology of China, Hefei, Anhui 230026, China}
\affiliation{Hefei National Laboratory, University of Science and Technology of China, Hefei 230088, China}
\author{Qian Niu}
\affiliation{International Centre for Quantum Design of Functional Materials, CAS Key Laboratory of Strongly-Coupled Quantum Matter Physics, and Department of Physics, University of Science and Technology of China, Hefei, Anhui 230026, China}

\date{\today}

\begin{abstract}
We theoretically investigate the engineering of two-dimensional second-order topological insulators with corner states by coupling two first-order topological insulators. We find that the interlayer coupling between two topological insulators with opposite topological invariants results in the formation of edge-state gaps, which are essential for the emergence of the corner states. Using the effective Hamiltonian framework, We elucidate that the formation of topological corner states requires either the preservation of symmetry in the crystal system or effective mass countersigns for neighboring edge states. Our proposed strategy for inducing corner state through interlayer coupling is versatile and applicable to both $\mathbb{Z}_2$ topological insulators and quantum anomalous Hall effects. We demonstrate this approach using several representative models including the seminal Kane-Mele model, the Bernevig-Hughes-Zhang model, and the Rashba graphene model to explicitly exhibit the formation of corner states via interlater coupling.
Moreover, we also observe that the stacking of the coupled $\mathbb{Z}_2$ topological insulating systems results in the formation of the time-reversal invariant three-dimensional second-order nodal ring semimetals. Remarkably, the three-dimensional system from the stacking of the Bernevig-Hughes-Zhang model can be transformed into second-order Dirac semimetals, characterized by one-dimensional hinge Fermi arcs. Our strategy of engineering second-order topological phases via simple interlayer coupling promises to advance the exploration of higher-order topological insulators in two-dimensional spinful systems.
\end{abstract}
\maketitle
\section{Introduction}
Topological materials have gathered considerable interest owing to their distinctive physical and mathematical principles and their potential applications in designing low-power electronic devices. Among these materials, topological insulators (TIs) stand out as novel quantum state materials distinguished by their insulating bulk states and topologically-protected conducting edge states in two-dimensional systems~\cite{Haldane1988, Kane2005, Kane2005a, Bernevig2006, Qiao2010, Qiao2011, Liu2008, Liu2023, Shen2024}, or surface states in three-dimensional systems~\cite{Qi2011, Hasan2010, Bansil2016, Ren2016}. Specifically, the presence of spin-helical edge states in $\mathbb{Z}_2$ TIs~\cite{Olsen2019, Choudhary2020, Kruthoff2017, Bradlyn2017, Po2017, Zhang2019, Vergniory2019, Tang2019}, protected by time-reversal symmetry, characterises topological invariants by $\mathbb{Z}_2$ numbers~\cite{Kane2005}. The simplest seminal models include the Kane-Mele model~\cite{Kane2005, Kane2005a} and the Bernevig-Hughes-Zhang model~\cite{Bernevig2006}, which are then extended to three dimensions~\cite{Fu2007, Roy2009}.
Another chiral edge state exists in the quantum anomalous Hall effect with broken time-reversal symmetry, which has also been extensively studied in different systems~\cite{Liu2008, Qiao2010, Chang2023, Nagaosa2010, Weng2015, Xu2011, Liu2016, He2018, Chang2023, Mei2024, Yu2010, Wu2014, Fang2014, Qiao2014, Wang2014, Xu2015, Sun2019, Wang2013, Qi2016, Hogl2020, Devakul2022, Chang2013, Chang2015, Chang2015a, Deng2020, Serlin2020}.

In recent years, both theoretical predictions and experimental observations have witnessed the emergence of zero-dimensional corner states in two-dimensional systems and one-dimensional hinge states in three-dimensional systems~\cite{Benalcazar2017, Benalcazar2017a, Li2020, Miert2018, Benalcazar2019, Schindler2019, Song2017, Schindler2018, Langbehn2017, Hsu2019, Schindler2018a, Peterson2018, Serra-Garcia2018}, leading to the exploration of more intriguing second-order TIs. While several theoretical proposals have suggested that materials such as black phosphorene~\cite{Ezawa2018}, twisted bilayer graphene at certain angles~\cite{Park2019}, and graphyne~\cite{Liu2019, Lee2020, Sheng2019} can exhibit the two-dimensional second-order TI, it is noteworthy that all these considerations have not invoked the electron spin degree of freedom.
In the year 2020, a close connection has been established between TIs and second-order TIs within two-dimensional spinful systems, mainly by introducing in-plane Zeeman fields in $\mathbb{Z}_2$ TIs~\cite{Ren2020}. The main approach to induce second-order topological corner states in other schemes is to apply in-plane Zeeman fields to break the time-reversal symmetry~\cite{Huang2022, Zhuang2022, Han2022, Miao2022, Miao2023, Miao2024, Chen2020}.

Besides, the second-order topological phases have also been studied in topological semimetals. In addition to the nodes (nodal lines) in the bulk, a distinctive characteristic of three-dimensional second-order semimetals is the presence of one-dimensional hinge Fermi arc states. For $k_z$-dependent two-dimensional ($k_x$-$k_y$) slices in three-dimensional second-order semimetals, a continuous second-order TI phase must exist. The corner states in these second-order TIs are connected to the hinge states in a three-dimensional topological semimetal~\cite{Lin2018, Ghorashi2020, Wang2020, Wang2020a}. This feature sets them apart from the surface Fermi arcs observed in conventional semimetals. However, so far, studies on higher-order semimetals have predominantly focused on spinless systems~\cite{Lin2018, Ezawa2018b, Wang2020, Wang2020a, Ghorashi2020, Xiong2023, Wei2021, Pu2023, Qiu2021, Ghorashi2021, Ghorashi2021a, Wang2022, Song2022}. In particular, there are no reports on the higher-order nodal line semimetals strictly in spinful system~\cite{Shao2021, Chen2022, Gao2023, Du2022}. To date, the utilization of stacking second-order TI to achieve higher-order nodal semimetals is mainly related to time-reversal breaking~\cite{Ezawa2018a, Zhang2021, Wieder2020, Szabo2020, Lee2022}. Therefore, two compelling questions naturally arise: (i) Is it possible to realize the two-dimensional second-order TIs without breaking the time-reversal symmetry in spinful systems? (ii) Is it possible to achieve three-dimensional time-reversal invariant second-order semimetals?

\begin{figure}[tbp]
  \centering
  \includegraphics[width=8.8cm,angle=0]{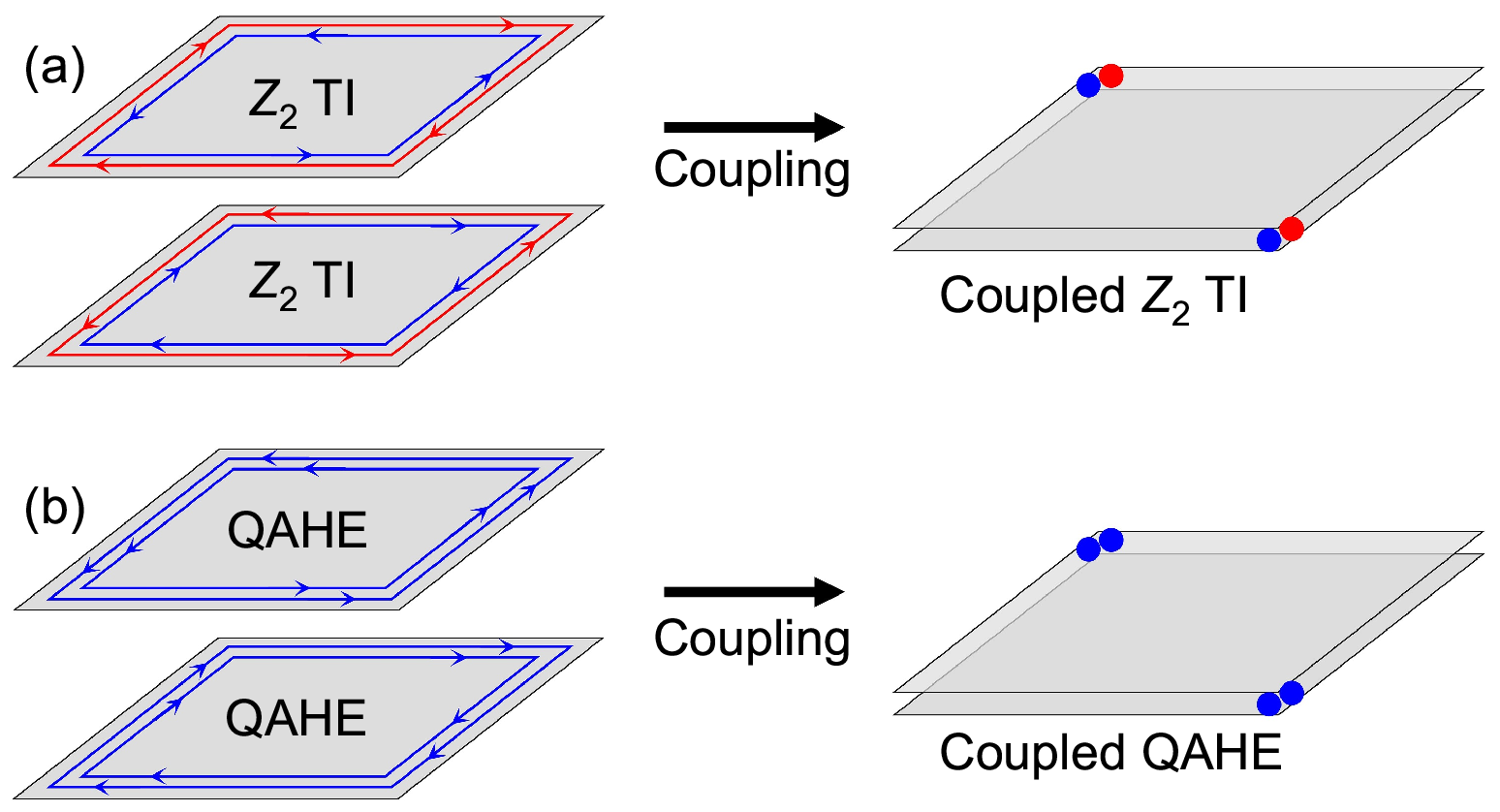}
  \caption{Schematic plot of coupled topological insulators. (a) Left: Two decoupled $\mathbb{Z}_2$ TI layers. TI layers with opposite spin-helical edge states, i.e., spin-up edge modes in red respectively propagate clockwise and counterclockwise in the top and bottom layers; while spin-down edge modes in blue respectively propagate counterclockwise and clockwise in the top and bottom layers.
  Right: Coupling destroys all kinds of edge modes and gives rise to two Kramer pairs of corner states denoted by red and blue dots.
  (b) Left: Two decoupled quantum anomalous Hall effect layers with opposite Chern numbers, i.e., the edge modes propagate clockwise/counterclockwise in the top/bottom layer. Right: Coupling destroys all kinds of edge modes, and gives rise to two corner states denoted by blue dots. QAHE denotes quantum anomalous Hall effect phase.
  }
  \label{fig1}
\end{figure}

In this article, we theoretically demonstrate that the second-order TI in two-dimensional spinful systems can be engineered by simply coupling two copies of $\mathbb{Z}_2$ TIs with opposite spin-helicities to produce Kramer pairs of corner states, as shown in Fig.~\ref{fig1}(a).
Before coupling, two decoupled $\mathbb{Z}_2$ TI layers with opposite spin-helical edge states, i.e., spin-up edge modes in red respectively propagate clockwise and counterclockwise in the top and bottom layers; while spin-down edge modes in blue respectively propagate counterclockwise and clockwise in the top and bottom layers. The introduction of interlayer coupling destroys the edge modes and gives rise to two Kramer pairs of corner states denoted by red and blue dots. Inspired by this finding, we further find that the coupling of two copies of quantum anomalous Hall effects with opposite chiralities (i.e., $\mathcal{C}=\pm N$) can also produce the second-order TI, i.e., corner states, as shown in Fig.~\ref{fig1}(b). Two decoupled quantum anomalous Hall effect layers with opposite Chern numbers, i.e., the edge modes propagate clockwise/counterclockwise in the top/bottom layer. The interlayer coupling destroys the edge modes and gives rise to two corner states denoted by blue dots. Since graphene is an ideal platform to realize the $\mathbb{Z}_2$ TI (by considering the intrinsic spin-orbit coupling)~\cite{Kane2005, Kane2005a}, quantum anomalous Hall effect (by considering Rashba spin-orbit coupling and exchange field simultaneously)~\cite{Qiao2010}. We mainly employ the graphene models to demonstrate the formation of corner states by coupling various topological phases. We also provide a minimal model to understand the physical origin of the formation of second-order TIs. Furthermore, we have utilized several other seminal model systems (e.g., the Bernevig-Hughes-Zhang model) to emphasize the universal features of our findings~\cite{Bernevig2006, Liu2008}.
A concise description of this work can be found in Ref.~\cite{Liu2024}.
In the end, we show that the stacking of the two-dimensional time-reversal invariant second-order TIs can lead to the formation of second-order nodal ring semimetal. By adjusting the orbital energy transfer, the nodal ring in a multilayered Bernevig-Hughes-Zhang model bulk is evolved into Dirac points, forming a second-order Dirac semimetal. The one-dimensional hinged Fermi arcs are connected to the nodal rings (nodes) of these three-dimensional topological semimetals as second-order topological properties.

The remaining of the paper is organized as follows. In Sec.~\ref{SectionII}, we consider various graphene-based topological systems to construct the bilayer hybrid structures with two opposite topological numbers to study the possibility of realizing second-order topological insulators (i.e., corner states). A low-energy effective model is provided to illustrate the physical origin of the corner states. In Sec.~\ref{SectionIII}, the topological phase diagrams and phase boundaries of the graphene-based corner states are presented. In Sec.~\ref{SectionIV}, to verify the universal formation mechanism of the corner states, we further consider several other representative topological systems. In Sec.~\ref{SectionV}, we stack time-reversal invariant second-order TIs along z-direction to obtain three-dimensional higher-order semimetals. In Sec.~\ref{SectionVI}, we summarise our findings of engineering second-order TIs via simply coupling two topological systems with opposite topological orders.

\section{Interlayer Coupling Induced Corner States in Graphene-Based Topological Systems}\label{SectionII}

It is well known that, due to the linear Dirac-type dispersion, graphene is an ideal platform to theoretically exhibit various topological phases, i.e., quantum spin-Hall effect (the earliest $\mathbb{Z}_2$ TI with the spin being a good quantum number) or Kane-Mele model from the intrinsic spin-orbit coupling~\cite{Kane2005, Kane2005a}, and quantum anomalous Hall effect with Rashba spin-orbit coupling and Zeeman field~\cite{Qiao2010}. The system Hamiltonian of our considered coupled graphene layers can then be described as:
\begin{align}
H =\left( \begin{matrix}
H_T&  \eta \\
\eta^*   &   H_B\\
\end{matrix} \right),
\label{eq4}
\end{align}
where $\eta$ is the coupling strength between the top and bottom graphene layers, and $H_{\rm T}$ and $H_{\rm B}$ are respectively related to the top and bottom graphene layers, whose corresponding tight-binding Hamiltonian in the presence of intrinsic spin-orbit coupling, Rashba spin-orbit coupling and Zeeman field can be expressed as following:
\begin{align}
H_{T/B}&=&-t\sum_{\left< i,j \right> ,\sigma}{c_{i\sigma}^{\dag}c_{j\sigma}}+i t_{\rm I} \sum_{\left< \left< i,j \right> \right> ,\sigma ,\sigma'}{\nu _{ij} c_{i\sigma}^{\dag}\left[ \mathbf{\hat{s}}_z \right] _{\sigma \sigma'}c_{j\sigma'}} \nonumber \\
&+& i t_{\rm R} \sum_{\left<  ij  \right> \sigma \sigma'}{ \mathbf{\hat{e}}_z \cdot ( \mathbf{s} \times \mathbf{d}_{ij}) c_{i\sigma}^{\dag} c_{j\sigma'}}+ \lambda \sum_{i \sigma }{c_{i\sigma}^{\dag} s_z c_{i\sigma}},
\label{eq5}
\end{align}
where $c_{i\sigma}$($c_{i\sigma}^{\dag}$) is the annihilation (creation) operator for an electron at site $i$ with $\sigma$ spin. $t$ is the hopping amplitude between the nearest neighboring sites. The second term is the intrinsic spin-orbit coupling between the next nearest neighbors, with $t_{\rm I}$ measuring the coupling strength. $\nu _{ij}=\pm1$ indicate counterclockwise and clockwise hopping paths from site $j$ to $i$, respectively. The third term is the Rashba spin-orbit coupling with coupling strength $t_{\rm R}$, and $\mathbf{d}_{ij}$ represents a unit vector pointing from site $j$ to site $i$. The last term corresponds to a uniform Zeeman field. Throughout this article, we measure the Fermi level, intrinsic and Rashba spin-orbit couplings, and Zeeman field in units of $t$.

\subsection{Coupled $\mathbb{Z}_2$ TIs of Kane-Mele models}\label{CoupledTI}
\begin{figure}
  \centering
  \includegraphics[width=8.8cm,angle=0]{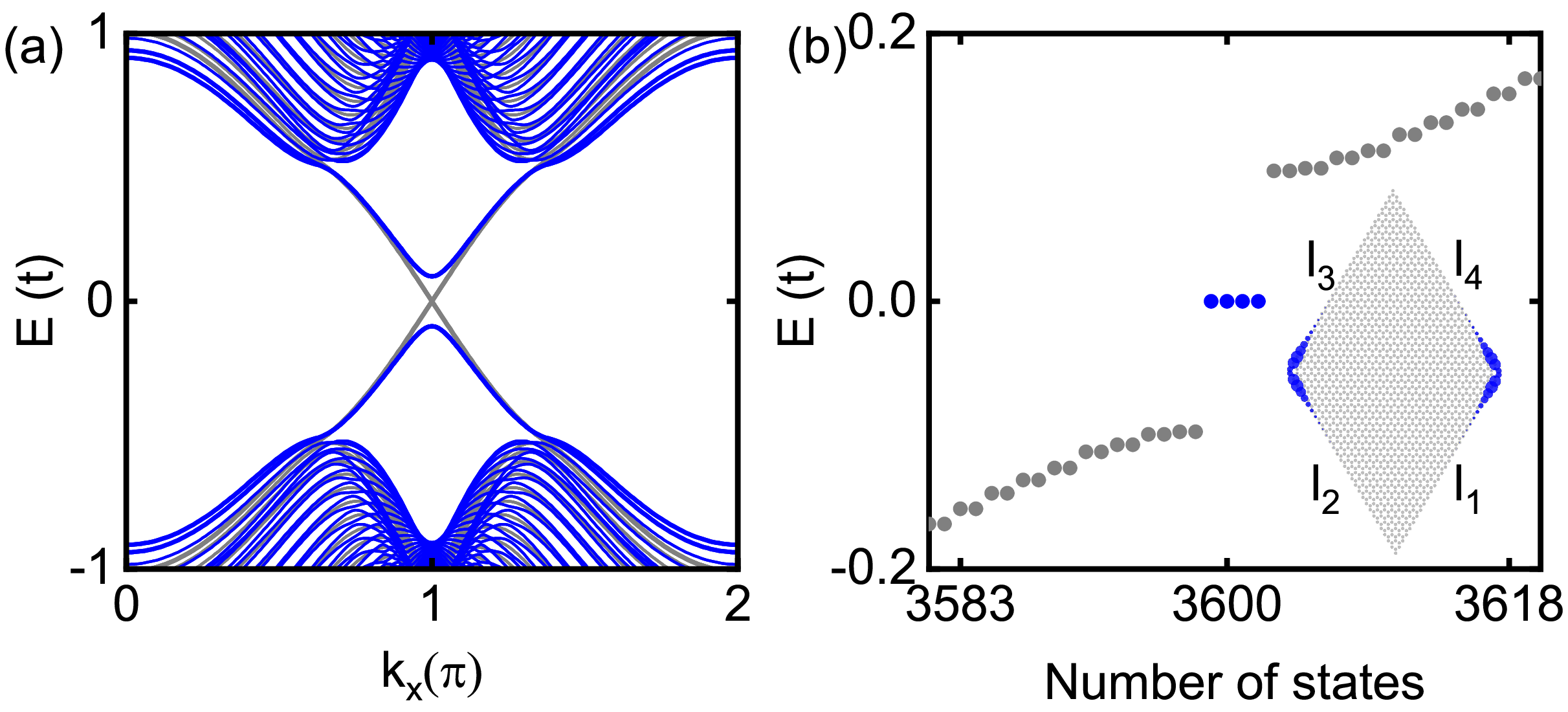}
  \caption{(a) Band structure of the Zigzag nanoribbon of the coupled Kane-Mele models. The energy bands with and without the interlayer coupling are shown in blue and gray, respectively. (b) Energy levels of diamond-shaped coupled nanoflakes. Inset: The probability distribution of the corner state. The parameters are set to be $t=1, t_{\rm I}^T= -t_{\rm I}^B= 0.1, \eta = 0.1$. The ribbon width is $N_y = 60a$, and the nanoflake size is $60a \times 60a$.
  }
  \label{fig2}
\end{figure}
Let us begin from the Kane-Mele system, as depicted in Fig.~\ref{fig1}(a), where the parameters are set to be $t=1, t_{\rm I}^T = -t_{\rm I}^B = 0.1, t_{\rm R}= \lambda= 0$. In our consideration, we choose the zigzag graphene ribbon width to be $N_y = 60a$, with $a$ being the lattice constant. When the two layers are isolated (i.e., $\eta = 0.0$), both exhibit the well-known spin-helical edge modes, i.e., edge modes with opposite spins counterpropagate along the same boundary. The only difference is that, at the top and bottom layers, the same spin-polarized edge mode propagates clockwise and counterclockwise, respectively, as shown by the gray line in Fig.~\ref{fig2}(a).

When the interlayer coupling $\eta = 0.1$ is considered, the blue line in Fig.~\ref{fig2}(a) illustrates the emergence of gapped helical edge states, suggesting the breaking of the gapless first-order $\mathbb{Z}_2$ TI. Remarkably, the system keeps the time-reversal symmetry, which distinguishes it from the gapped edge states induced by the in-plane magnetization~\cite{Ren2020}. To determine the topological nature of the opened band gap, we plot the energy level of a nanoflake with zigzag boundaries, as depicted in Fig.~\ref{fig2}(b). Specifically, we analyze a typical diamond-shaped flake with a width of $L_n = 60a$. Interestingly, within the edge-state gap, two pairs of Kramer-degenerate zero-energy states arise, denoted by blue dots. The probability distributions of the wave functions for these in-gap states at half-filling are illustrated in the inset of Fig.~\ref{fig2}(b). In the diamond-shaped sample, the zero-energy in-gap state with fractional charge $e/2$ is primarily localized at the two obtuse corners. This phenomenon directly indicates the formation of a second-order TI, which is characterized by corner states in two-dimensional systems.

Now, we exploit the formation of edge state gaps from the low-energy effective Hamiltonian. The combination of two $2 \times 2$ modified Dirac models can generate an effective model for the quantum spin-Hall effect~\cite{Shen2017}. Under the time-reversal operator $\mathcal{T} = i s_y \mathcal{K}$,
\begin{align}
 \textbf{k} \longrightarrow -\textbf{k}, \sigma_{i} \longrightarrow -\sigma_{i},
 \label{eq20}
\end{align}
therefore,
\begin{align}
 \mathcal{T} \textbf{d(k)} \cdot \sigma \mathcal{T}^{-1} = -\textbf{d}(-\textbf{k}) \cdot \sigma,
 \label{eq21}
\end{align}
where $\textbf{d(k)} \cdot \sigma$ and $-\textbf{d}(-\textbf{k}) \cdot \sigma$ represent the spin-up and spin-down sectors, respectively. In the representative report~\cite{Ren2020} of realizing second-order TI from introducing the in-plane magnetization to a $\mathbb{Z}_2$ TI, on the basis of \{$\psi_{\uparrow}, \psi_{\downarrow}$\}, its effective Hamiltonian can be written as:
\begin{align}
   h_{0}=\left[ \begin{matrix}
 \textbf{d(k)} \cdot \sigma &  B_x \\
 B_x  &  -\textbf{d}(-\textbf{k}) \cdot \sigma \\
\end{matrix} \right],
 \label{eq1}
\end{align}
where $B_x$ is the in-plane Zeeman field, which can be seen as a coupling/scattering between spin-up and spin-down edge states propagating along opposite directions, leading to the edge gap opening.

In our coupled system, the same spin-polarized edge states propagate along opposite directions at different layers. The corresponding effective Hamiltonian on the basis of \{$\psi_{T\uparrow}, \psi_{T\downarrow}, \psi_{B\uparrow}, \psi_{B\downarrow}$\} can be written as:
\begin{eqnarray}
 h_{1}  =\left[ \begin{matrix}
\textbf{d}_T(\textbf{k}) \cdot \sigma &  0 &  \eta  &  0 \\
 0  &  -\textbf{d}_T(-\textbf{k}) \cdot \sigma &  0 &  \eta\\
 \eta^*  &  0 &  \textbf{d}_B(-\textbf{k}) \cdot \sigma  &  0 \\
 0 &  \eta^* &  0  &  -\textbf{d}_B(\textbf{k}) \cdot \sigma\\
\end{matrix} \right]\nonumber,
 \label{eq-eff2}
\end{eqnarray}
where $\textbf{d}_{\rm T}$ and $\textbf{d}_{\rm B}$ represent the top and bottom $\mathbb{Z}_2$ TIs, respectively. $\eta$ denotes the interlayer coupling, acting on edge states with different propagation directions. Since spin is a good quantum number in the Kane-Mele model, one can consider the spin-up sectors separately. Thus, the Hamiltonian on the basis of \{$\psi_{T\uparrow}, \psi_{B\uparrow}$\} can be written as:
\begin{eqnarray}
   h_{2}=\left[
   \begin{matrix}
 \textbf{d}_T\textbf{(k)} \cdot \sigma &  \eta \\
 \eta^*  &  \textbf{d}_B(-\textbf{k}) \cdot \sigma \\
\end{matrix} \right].
 \label{eq-eff3}
\end{eqnarray}
One can see that Eqs.~(\ref{eq1}) and ~(\ref{eq-eff3}) share the same form, with only the interlayer degree of freedom and coupling intensity being replaced by the spin and in-plane Zeeman field. The sector of spin-down is similar. Compared with the in-plane magnetization-induced second-order TI, our interlayer coupling-induced one does not break the time-reversal symmetry.

To intuitively grasp the topological origin of corner states, we investigate edge states in the coupled graphene system. Specifically, when considering a two-layer Kane-Mele model devoid of interlayer coupling, boundary states at zigzag boundaries can be analytically resolved. In the inset of Fig.~\ref{fig2}(b), the four boundaries of the diamond-shaped sample are labeled as $l_1, l_2, l_3,$ and $l_4$, and the associated spin-helical edge states (spinor part) are provided by~\cite{Tan2022}:
\begin{widetext}
\begin{eqnarray}
{\left| {{\chi _ {T,\uparrow} }} \right\rangle ^{l_{1}/l_{2}}} &=
\frac{1}{{\sqrt {1 + 1/\mu } }}{\left[ {\begin{array}{*{20}{c}}
1\\
0
\end{array}} \right]_\tau} \otimes{\left[ {\begin{array}{*{20}{c}}
1\\
{ - \frac{i}{\sqrt \mu}  }
\end{array}} \right]_\sigma } \otimes {\left[ {\begin{array}{*{20}{l}}
1\\
0
\end{array}} \right]_s},  \qquad
{\left| {{\chi _ {T,\downarrow} }} \right\rangle ^{l_{1}/l_{2}}} =
\frac{1}{{\sqrt {1 + 1/\mu } }}{\left[ {\begin{array}{*{20}{c}}
1\\
0
\end{array}} \right]_\tau} \otimes{\left[ {\begin{array}{*{20}{c}}
1\\
{ \frac{i}{\sqrt \mu}  }
\end{array}} \right]_\sigma } \otimes {\left[ {\begin{array}{*{20}{l}}
0\\
1
\end{array}} \right]_s}, \nonumber\\
{\left| {{\chi _ {B,\uparrow} }} \right\rangle ^{l_{1}/l_{2}}} &=
\frac{1}{{\sqrt {1 + 1/\mu } }}{\left[ {\begin{array}{*{20}{c}}
0\\
1
\end{array}} \right]_\tau} \otimes{\left[ {\begin{array}{*{20}{c}}
1\\
{  \frac{i}{\sqrt \mu}  }
\end{array}} \right]_\sigma } \otimes {\left[ {\begin{array}{*{20}{l}}
1\\
0
\end{array}} \right]_s},  \qquad
{\left| {{\chi _ {B,\downarrow} }} \right\rangle ^{l_{1}/l_{2}}} =
\frac{1}{{\sqrt {1 + 1/\mu } }}{\left[ {\begin{array}{*{20}{c}}
0\\
1
\end{array}} \right]_\tau} \otimes{\left[ {\begin{array}{*{20}{c}}
1\\
{ - \frac{i}{\sqrt \mu}  }
\end{array}} \right]_\sigma } \otimes {\left[ {\begin{array}{*{20}{l}}
0\\
1
\end{array}} \right]_s}, \nonumber
\end{eqnarray}
\begin{eqnarray}
{\left| {{\chi _ {T,\uparrow} }} \right\rangle ^{l_{3}/l_{4}}} &=
\frac{1}{{\sqrt {1 + \mu } }}{\left[ {\begin{array}{*{20}{c}}
1\\
0
\end{array}} \right]_\tau} \otimes{\left[ {\begin{array}{*{20}{c}}
1\\
{ - i\sqrt \mu  }
\end{array}} \right]_\sigma } \otimes {\left[ {\begin{array}{*{20}{l}}
1\\
0
\end{array}} \right]_s}, \qquad
{\left| {{\chi _ {T,\downarrow} }} \right\rangle ^{l_{3}/l_{4}}} =
\frac{1}{{\sqrt {1 + \mu } }}{\left[ {\begin{array}{*{20}{c}}
1\\
0
\end{array}} \right]_\tau} \otimes{\left[ {\begin{array}{*{20}{c}}
1\\
{ i\sqrt \mu  }
\end{array}} \right]_\sigma } \otimes {\left[ {\begin{array}{*{20}{l}}
0\\
1
\end{array}} \right]_s}, \nonumber\\
{\left| {{\chi _ {B,\uparrow} }} \right\rangle ^{l_{3}/l_{4}}} &=
\frac{1}{{\sqrt {1 + \mu } }}{\left[ {\begin{array}{*{20}{c}}
0\\
1
\end{array}} \right]_\tau} \otimes{\left[ {\begin{array}{*{20}{c}}
1\\
{  i\sqrt \mu  }
\end{array}} \right]_\sigma } \otimes {\left[ {\begin{array}{*{20}{l}}
1\\
0
\end{array}} \right]_s}, \qquad
{\left| {{\chi _ {B,\downarrow} }} \right\rangle ^{l_{3}/l_{4}}} =
\frac{1}{{\sqrt {1 + \mu } }}{\left[ {\begin{array}{*{20}{c}}
0\\
1
\end{array}} \right]_\tau} \otimes{\left[ {\begin{array}{*{20}{c}}
1\\
{ - i\sqrt \mu  }
\end{array}} \right]_\sigma } \otimes {\left[ {\begin{array}{*{20}{l}}
0\\
1
\end{array}} \right]_s}, \nonumber\\
\end{eqnarray}
with $\mu  = \left( {1 + \frac{{{t^2}}}{{8 t_{\rm I}^2}}} \right) - \sqrt {{{\left( {1 + \frac{{{t^2}}}{{8 t_{\rm I}^2}}} \right)}^2} - 1}$,  $\tau$, $\sigma$, and $s$ repersent interlayer, sublattice and spin degrees of freedom, respectively .
The effective mass term ${\bf M}_{\eta}^{l_{1}}$  of the edge $l_{1}$ can be obtained by projecting the interlayer coupling term $H_\eta = \eta {\tau_x}{\sigma _0}{s_0}$  onto the subspace spanned by $ | {\chi _ {T,\uparrow} ^{l_{1}}}  \rangle$, $ | {\chi _ {T,\downarrow} ^{l_{1}}}  \rangle$, $ | {\chi _ {B,\uparrow} ^{l_{1}}}  \rangle$ and $ | {\chi _ {B,\downarrow} ^{l_{1}}}  \rangle$.
Then, we can obtain the effective mass of the $l_{1}$ boundary,
\begin{eqnarray}
{{\bf M}_{\eta}^{l_{1}}} &= & {\eta}\left( {\begin{array}{*{20}{c}}
0&0&{\langle \chi _ {T,\uparrow} ^{l_1}|{\tau_x}{\sigma _0}{s_0} | {\chi _ {B,\uparrow} ^{l_1}} \rangle }&0\\
0&0&0&{\langle \chi _ {T,\downarrow} ^{l_1}|{\tau_x}{\sigma _0}{s_0} | {\chi _ {B,\downarrow} ^{l_1}} \rangle }\\
{\langle \chi _ {B,\uparrow} ^{l_1}|{\tau_x}{\sigma _0}{s_0} | {\chi _ {T,\uparrow} ^{l_1}}  \rangle }&0&0&0\\
0&{\langle \chi _ {B,\downarrow} ^{l_1}|{\tau_x}{\sigma _0}{s_0} | {\chi _ {T,\downarrow} ^{l_1}}  \rangle }&0&0
\end{array}} \right)\nonumber\\
& = & {\eta}\frac{{{\rm{|}}\mu {\rm{|}}-1}}{{{\rm{|1 + }}\mu {\rm{|}}}}\left( {\begin{array}{*{20}{c}}
0&0&1&0\\
0&0&0&1\\
1&0&0&0\\
0&1&0&0
\end{array}} \right).
\end{eqnarray}

Similarly, the effective mass for the $l_3$ boundary can be expressed as:
\begin{eqnarray}
{{\bf M}_{\eta}^{l_{3}}} &= & {\eta}\left( {\begin{array}{*{20}{c}}
0&0&{\langle \chi _ {T,\uparrow} ^{l_3}|{\tau_x}{\sigma _0}{s_0} | {\chi _ {B,\uparrow} ^{l_3}}  \rangle }&0\\
0&0&0&{\langle \chi _ {T,\downarrow} ^{l_3}|{\tau_x}{\sigma _0}{s_0} | {\chi _ {B,\downarrow} ^{l_3}}  \rangle }\\
{\langle \chi _ {B,\uparrow} ^{l_3}|{\tau_x}{\sigma _0}{s_0} | {\chi _ {T,\uparrow} ^{l_3}}  \rangle }&0&0&0\\
0&{\langle \chi _ {B,\downarrow} ^{l_3}|{\tau_x}{\sigma _0}{s_0} | {\chi _ {T,\downarrow} ^{l_3}}  \rangle }&0&0
\end{array}} \right)\nonumber\\
& = & {\eta}\frac{{1-{\rm{|}}\mu {\rm{|}}}}{{{\rm{|1 + }}\mu {\rm{|}}}}\left( {\begin{array}{*{20}{c}}
0&0&1&0\\
0&0&0&1\\
1&0&0&0\\
0&1&0&0
\end{array}} \right).
\end{eqnarray}
\end{widetext}

Since the effective mass term for the $l_1 (l_3)$ boundary is exactly the same as that for the $l_2 (l_4)$ boundary, there are no corner states at the acute corners. Surprisingly, it is evident that the effective masses of the boundaries $l_1 (l_2)$ and $l_4 (l_3)$ possess opposite signs.
According to the Jackiw-Rebbi theory~\cite{Jackiw1976}, this configuration forms a wall of mass domains at the obtuse corners, therefore facilitating the emergence of topologically protected corner states [see in the inset of Fig.~\ref{fig2}(b)].
\begin{figure}
  \centering
  \includegraphics[width=8.6cm,angle=0]{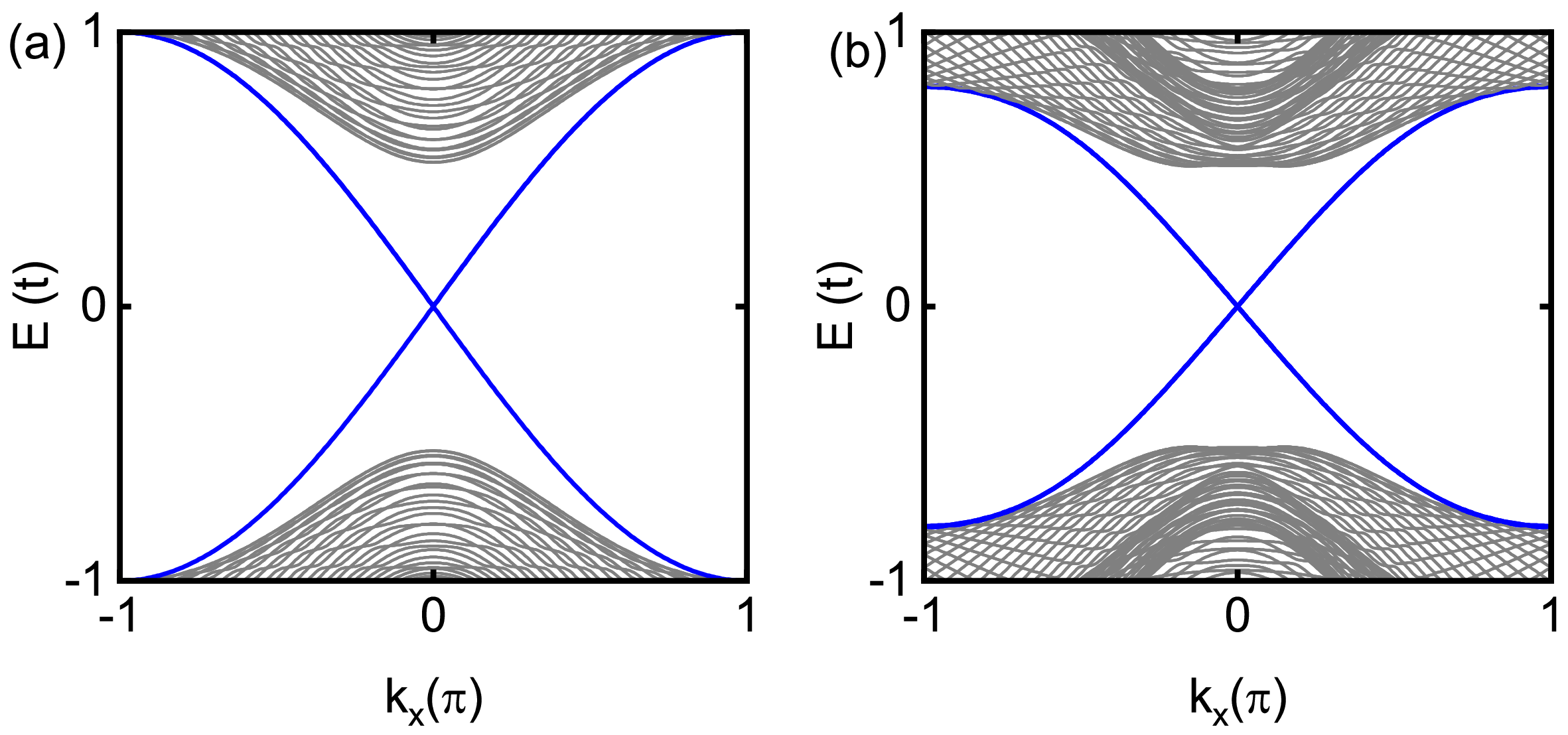}
  \caption{(a) Band spectra for the armchair nanoribbon of the decoupled Kane-Mele models. (b) Band spectra for the armchair ribbon of the coupled Kane-Mele models with an interlayer coupling of $\eta = 0.1$. Other parameters are set to be: $t = 1, t_{\rm I}^{\rm T} = -t_{\rm I}^{\rm B} = 0.1, t_{\rm R}= 0, \lambda = 0$.
  }
  \label{fig3}
\end{figure}

Due to the peculiarities of the $K$ and $K'$ valley positions in the Brillouin zone of graphene, the energy band folding induced by interlayer coupling is inequivalent at the zigzag and armchair boundaries and is similar to the in-plane Zeeman field in Kane-Mele model~\cite{Ren2020}. This simple folding of the energy bands does not change the topology of the armchair boundary, and therefore the edge states remain gapless. Since this system is no longer protected from the time-reversal symmetry due to the presence of two-pair spin-helical edge modes, it becomes a weak TI with $\mathbb{Z}_2 =0$. In Fig.~\ref{fig3}, we present the band structures of the armchair ribbon in the absence and presence of interlayer coupling, i.e., $\eta = 0.0$ (a) and 0.1 (b). Other parameters are set to be $t=1, t_{\rm I}^T = -t_{\rm I}^B = 0.1$. It is noteworthy that the interlayer coupling does not destroy the gapless edge states. However, it is essential to emphasize that these gapless edge states exclusively manifest on the armchair boundary. Therefore, it is necessary to utilize zigzag-boundary nanoflakes to engineer higher-order TIs in honeycomb lattice systems.

\begin{figure}
  \centering
  \includegraphics[width=8.8cm,angle=0]{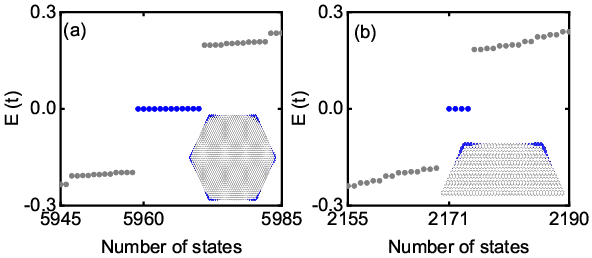}
  \caption{Energy levels for (a) hexagonal-shaped and (b) trapezoidal-shaped nanoflakes with zigzag boundaries of the coupled TIs. Inset: The probability distribution of the corner state. The parameters are set to be $t=1, t_{\rm I}^T= -t_{\rm I}^B= 0.1, t_{\rm R}= 0, \eta = 0.2$.
  }
  \label{figtx}
\end{figure}

It is well-known that only 60-degree and 120-degree angles exist in graphene nanoflakes consisting of zigzag boundaries. According to the edge-state Dirac effective mass analysis: when two zigzag boundaries form a 60-degree angle, no mass-domain wall is formed; when they form a 120-degree angle, mass-domain walls are formed to support the corner states. To verify the origin of the corner states, we further analyze the electronic properties of hexagonal and trapezoidal nanoflakes with zigzag boundaries, in Fig.~\ref{figtx}. In Fig.~\ref{figtx}(a), we show the energy level and wave function distributions of the hexagonal nanoflake. As expected, twelve in-gap corner states are distributed at the six corners of the nanoflake due to the time-reversal symmetry and the Dirac mass domain wall (see inset). In Fig.~\ref{figtx}(b), we show the energy level and wave function distributions for the trapezoidal nanoflake with 60-degree and 120-degree corners. One can see that the four in-gap corner states are distributed at the two 120-degree corners, with no corner states at the 60-degree corners. These observations justify our analysis of the origin of the corner state and provide valuable guidance for the possible experimental realization.

\begin{figure}
  \centering
  \includegraphics[width=8.8cm,angle=0]{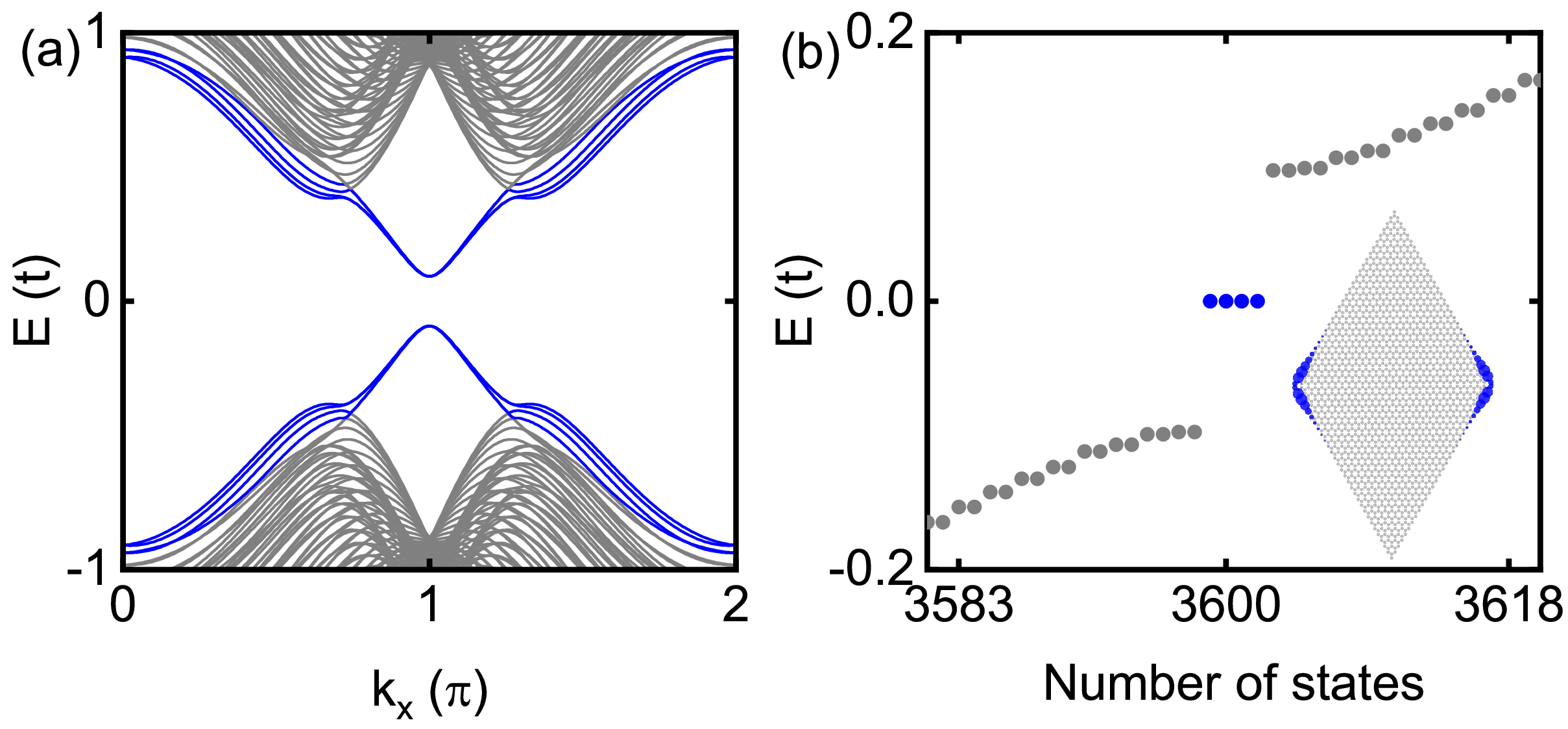}
  \caption{(a) Band structures of the zigzag nanoribbon for the coupled Kane-Mele model in the presence of Rashba spin-orbit coupling.
  (b) Energy levels of diamond-shaped nanoflakes. Inset: Probability distribution of the corner state. The parameters are set to be $t=1, t_{\rm I}^T= -t_{\rm I}^B= 0.1, t_{\rm R}= 0.05, \eta = 0.1$. The ribbon width is $N_y = 60a$, and the nanoflake size is $60a \times 60a$.
  }
  \label{figra}
\end{figure}

The Kane-Mele model serves as a valuable tool to investigate the fundamental physics of topological insulators. However, it experiences challenges due to the inherently weak intrinsic spin-orbit coupling. Nonetheless, theoretical advancements have been made in this regard. Drawing parallels with graphene, atomic layers of other group IV elements, such as silicene, germanene, and stanene~\cite{Liu2011, Ezawa2012, Ezawa2015, Molle2017}, demonstrate significantly stronger intrinsic spin-orbit coupling. Fortunately, all these systems can be properly characterized by the modified Kane-Mele model with Rashba spin-orbit coupling. Therefore, it is necessary to discuss the influence of Rashba spin-orbit coupling on corner states. In Fig.~\ref{figra}(a), we show the band structure of the zigzag ribbons with Rashba spin-orbit coupling $t_{\rm R}= 0.05$, with other parameters being the same as those in Fig.~\ref{fig2}. The Rashba spin-orbit coupling leads to band splitting of both bulk and ribbons. In Fig.~\ref{figra}(b), we plot the energy levels and wave function distributions for the diamond-shaped nanoflake.
One can see that the position and number of corner states are unchanged. Therefore, the higher-order topological corner states induced in the coupled Kane-Mele models are robust against the external Rashba spin-orbit coupling.

\subsection{Coupling graphene-based quantum anomalous Hall effect}
Inspired by the aforementioned findings, a natural question arises: if we replace the time-reversal symmetry protected $\mathbb{Z}_2$ TI by the time-reversal symmetry broken quantum anomalous Hall effect, whether the interlayer coupling between two quantum anomalous Hall effects with opposite Chern numbers can still be able to produce a second-order TI?

\begin{figure}
  \centering
  \includegraphics[width=8.8cm,angle=0]{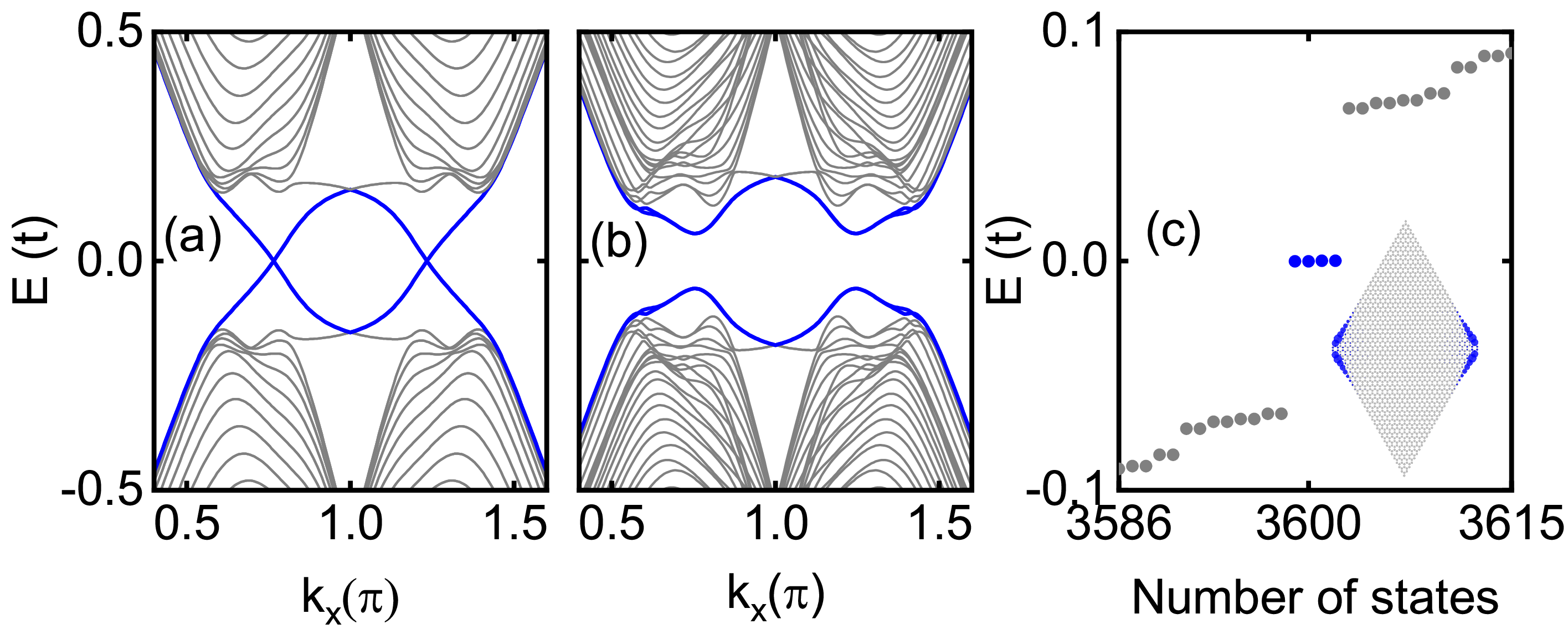}
  \caption{(a)-(b) Band structures of the zigzag nanoribbons for the (a) decoupled and (b) coupled graphene-based quantum anomalous Hall effects with Rashba spin-orbit coupling and Zeeman field. The interlayer coupling strength is set to be $\eta = 0.1$. The blue lines denote the chiral edge states. (c) Energy levels of diamond-shaped coupled nanoflakes with the parameters being the same as those in (b). Corner states are highlighted in blue. The probability distribution of the corner state is plotted in the inset. Other system parameters are chosen to be $t = 1, t_{\rm R}= 0.2, \lambda_T= - \lambda_B= 0.2$, and the nanoribbon width $N_y = 60a$ for (a) and (b). The nanoflake size $60a \times 60a$ for (c).
  }
  \label{fig4}
\end{figure}

To answer this question, we use the graphene model with Rashba spin-orbit coupling and Zeeman field to produce the quantum anomalous Hall effect. In our considerations, we set the system parameters to be $t = 1, t_{\rm R}= 0.2, \lambda_T= -\lambda_B= 0.2, t_{\rm I} = 0$, with the zigzag ribbon width of $N_y = 60a$. We first calculate the band structure of the zigzag boundary when the interlayer coupling is vanishing, as plotted in Fig.~\ref{fig4}(a). One can see that two-pair chiral edge states appear near the $K$ and $K'$ points. Since $\lambda_T= -\lambda_B$, the chiral edge modes counterpropagate at the top and bottom graphene layers. When the interlayer coupling $\eta = 0.1$ is introduced, the resulting band structure can be seen in Fig.~\ref{fig4}(b), where an edge gap opens. The emergence of gapped edge states is necessary for the formation of second-order TI.

To ascertain whether the opened edge gap is topologically trivial or nontrivial, we plot the energy levels of the diamond-shaped nanoflakes with zigzag boundaries in Fig.~\ref{fig4}(c), with a flake length of $L_n = 60a$. One can see that four zero-energy in-gap states are formed, indicated by the blue dots. Furthermore, the probability distribution of the wave function at half-filling is highlighted in the inset of Fig.~\ref{fig4}(c). Since the chosen sample was quantum anomalous Hall effects with Chern numbers of $\mathcal{C}=2/-2$, four corner states with a fractional charge of $e/2$ for each state are primarily localized at the two obtuse angles of the diamond-shaped nanoflake. These findings strongly suggest that the interlayer coupling can also drive the hybrid structure of quantum anomalous Hall effects with opposite Chern numbers to be a second-order TI. It is noteworthy that, even without the protection of time-reversal symmetry, the presence of degenerate corner states at individual corners is guaranteed. Based on the analysis of the corner state origins, the coupled graphene-based quantum anomalous Hall effect model can also be utilized to induce corner states using other configurations of zigzag-boundary nanosheets, such as trapezoidal and hexagonal nanoflakes.

\section{phase diagram for second-order topological insulators}\label{SectionIII}

\begin{figure*}[htbp]
  \centering
  \includegraphics[width=18cm,angle=0]{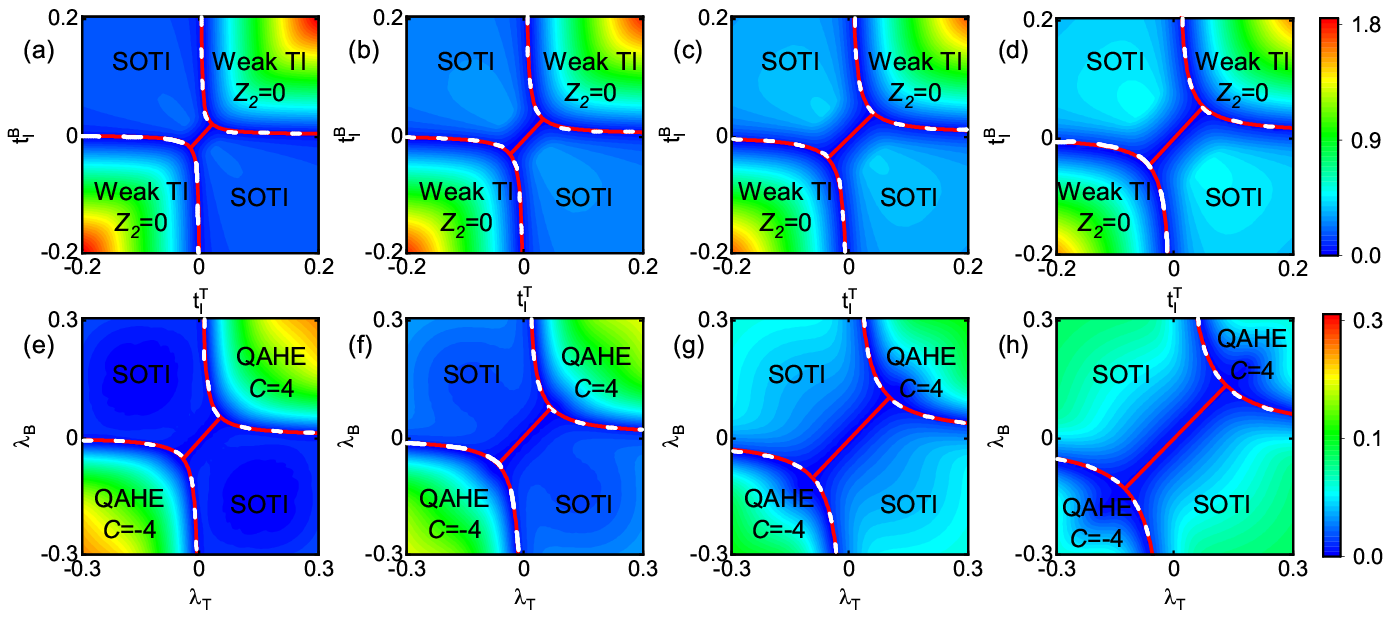}
  \caption{(a)-(d) Topological phase diagram of the coupled Kane-Mele models as functions of $t_{\rm I}^T$ and $t_{\rm I}^B$ at different interlayer coupling strengths $\eta=0.1, 0.15, 0.2$, and 0.25. Other parameters are set to be $t=1, t_{\rm R}=0, \lambda_{T,B}=0$. The dashed line is the phase boundary analytically obtained from Eq.~(\ref{eq8}) between the second-order TI and weak TI phases.
  (e)-(h) Topological phase diagram of the coupled quantum anomalous Hall effects with $\mathcal{C}=\pm2$ as functions of $\lambda_T$ and $\lambda_B$ at different interlayer coupling strengths $\eta=0.05, 0.07, 0.1,$ and $0.13$, with $t_R = 0.2$. Other parameters are set to be $t=1, t_{\rm I}^{T,B}=0$. The dashed line is the phase boundary obtained from Eq.~(\ref{eq9}) between the second-order TI and $\mathcal{C}= \pm 4$ quantum anomalous Hall phase. Color measures the bulk gap amplitude in the first-order TI phases and the edge gap amplitude in the second-order TI phases.
  QAHE and SOTI represent the quantum anomalous Hall effect phase and the second-order TI phase, respectively.
  }
  \label{fig5}
\end{figure*}

In the above discussions, we focus on the specific scenarios, where the determining parameters are exactly opposite, e.g., $t_{\rm I}^T = - t_{\rm I}^B$ for coupled $\mathbb{Z}_2$ TIs, and $\lambda_T = -\lambda_B$ for coupled quantum anomalous Hall systems. Intuitively, for the coupled $\mathbb{Z}_2$ TI system, one may expect that when the signs of $t_{\rm I}^T$ and $t_{\rm I}^B$ are identical, the coupling of two $\mathbb{Z}_2$ TIs results in the formation of the weak TI, which exhibits two (even) pairs of gapless spin-helical edge modes but $\mathbb{Z}_2 = 0$; when the signs of $t_{\rm I}^T$ and $t_{\rm I}^B$ are different, the coupling leads to the formation of second-order TI. For the coupled quantum anomalous Hall effects, one may expect that when the signs of $\lambda_T$ and $\lambda_B$ are identical, the coupling results in the formation of a high-Chern-number quantum anomalous Hall effect, which exhibits four gapless chiral edge modes with $\mathcal{C} = 4$; when the signs of $\lambda_T$ and $\lambda_B$ are different, the coupling leads to the formation of second-order TI.

To verify this hypothesis, we numerically determine the topological phase diagrams for the coupled Kane-Mele models within the parameter spaces of $t_{\rm I}^T$ and $t_{\rm I}^B$ at various coupling strengths (e.g., $\eta=0.1, 0.15, 0.2$, and 0.25) in Figs.~\ref{fig5}(a)-~\ref{fig5}(d). In addition, Figures~\ref{fig5}(e)-~\ref{fig5}(h) show the topological phase diagrams for the coupled quantum anomalous Hall system within the parameter spaces of $\lambda_T$ and $\lambda_B$ at various coupling strengths (e.g., $\eta=0.05, 0.07, 0.1,$ and $0.13$). One can find that the regimes before and after gap closing are characterized by different topological phases. In Figs.~\ref{fig5}(a)-~\ref{fig5}(d), regimes II and IV correspond to the second-order TI phases, while regimes I and III are the weak TI phases. In Figs.~\ref{fig5}(e)-~\ref{fig5}(h), one can see that regimes II and IV are the second-order TI phases, while regimes I and III are the $\mathcal{C}=4$ high Chern number quantum anomalous Hall phases. Furthermore, by comparing the phase diagrams from Fig.~\ref{fig5}(a) to Fig.~\ref{fig5}(d) (or from Fig.~\ref{fig5}(e) to Fig.~\ref{fig5}(h)), one can obtain that the size of the edge state gap is closely related to the interlayer coupling strength.

At the critical point of the topological phase transition, the gap-closing condition allows us to obtain an analytic expression of the phase transition boundary from the low-energy continuum Hamiltonian. By expanding the tight-binding Hamiltonian Eq. (\ref{eq4}) at the vicinity of $K$, one can get a eight-band low-energy Hamiltonian:
\begin{align}
H=\left( \begin{matrix}
H_T &  H_\eta \\
H_\eta^* &  H_B\\
\end{matrix} \right),
\label{eq6}
\end{align}
where, $H_T, H_B,$ and $H_\eta$ are written as:
\begin{eqnarray}
H_T = &&v (\sigma_x k_x +\sigma_y k_y) + \frac {M_I^T} {2} \sigma_z s_z + M_R (\sigma_x s_y - \sigma_y s_x) \nonumber \\
&& + \lambda_T s_z \textbf{1}_\sigma, \\
H_B = && v (\sigma_x k_x +\sigma_y k_y) + \frac {M_I^B} {2} \sigma_z s_z + M_R (\sigma_x s_y - \sigma_y s_x) \nonumber \\
&& + \lambda_B s_z \textbf{1}_\sigma,\\
H_\eta =&& \eta \textbf{1}_s \textbf{1}_\sigma,
\label{eq7}
\end{eqnarray}
where $\sigma$ and $s$ are Pauli matrices representing the $A, B$ sublattice, and spin degrees of freedom, respectively. The Fermi velocity, Rashba coupling, and Intrinsic spin-orbit coupling are given, respectively, by $v= 3 t a /2, M_R = 3 t_R,$ and $M_I^T = 3\sqrt{3} t_I^T, M_I^B = 3\sqrt{3} t_I^B$.

The low-energy Hamiltonian at $k = 0$ corresponds to the gap closing condition with the energy eigenvalues being $\varepsilon=0$. For the coupled Kane-Mele model systems, by setting $t_R = 0, \lambda_{T, B} = 0$, the topological phase transition boundary is solved to satisfy the following relation:
\begin{align}
t_{\rm I}^T t_{\rm I}^B =\eta^2 /27.
\label{eq8}
\end{align}
For the coupled quantum anomalous Hall effect model, by setting $t_I^T = t_I^B= 0$, the topological phase transition boundary is solved to satisfy the following relation:
\begin{align}
\lambda_T \lambda_B =\eta^2.
\label{eq9}
\end{align}
In Figs.~\ref{fig5}(a)-\ref{fig5}(d), we plot the phase boundaries according to Eq.~(\ref{eq8}) with dashed lines separating the second-order TI from the weak TI; while in Figs.~\ref{fig5}(e)-\ref{fig5}(h), we plot the phase boundaries according to Eq.~(\ref{eq9}) with dashed lines separating the second-order TI from the quantum anomalous Hall phase with Chern number of $\mathcal{C}=\pm 4$. One can find that the phase boundaries from our analytic expressions completely overlap with those from the band structure calculation based on the tight-binding model Hamiltonian. However, in Fig.~\ref{fig5}, in addition to the phase boundary described by the curves corresponding to Eq.~(\ref{eq8}) and Eq.~(\ref{eq9}), there is another phase boundary connecting the two curves. This additional boundary arises from the small bulk gap in Kane-Mele-type graphene with weak spin-orbit coupling. The interlayer coupling induces a cleavage between the bands of the top and bottom layers, resulting in the closure of the energy bands. As can be seen in Figs.~\ref{fig5}(a)-\ref{fig5}(d), the straight-line phase boundary becomes progressively longer as the interlayer coupling increases. For the quantum anomalous Hall effect with a weak Zeeman field, the bulk gap is also very small. Similarly, as the interlayer coupling strength increases, the straight-line phase boundary becomes longer and even crosses two curves, as displayed in Figs.~\ref{fig5}(e)-\ref{fig5}(h). Importantly, this extended phase boundary does not affect the weak TI, quantum anomalous Hall effect, or the second-order TI phase.

\begin{figure}[htbp]
  \centering
  \includegraphics[width=8.8cm,angle=0]{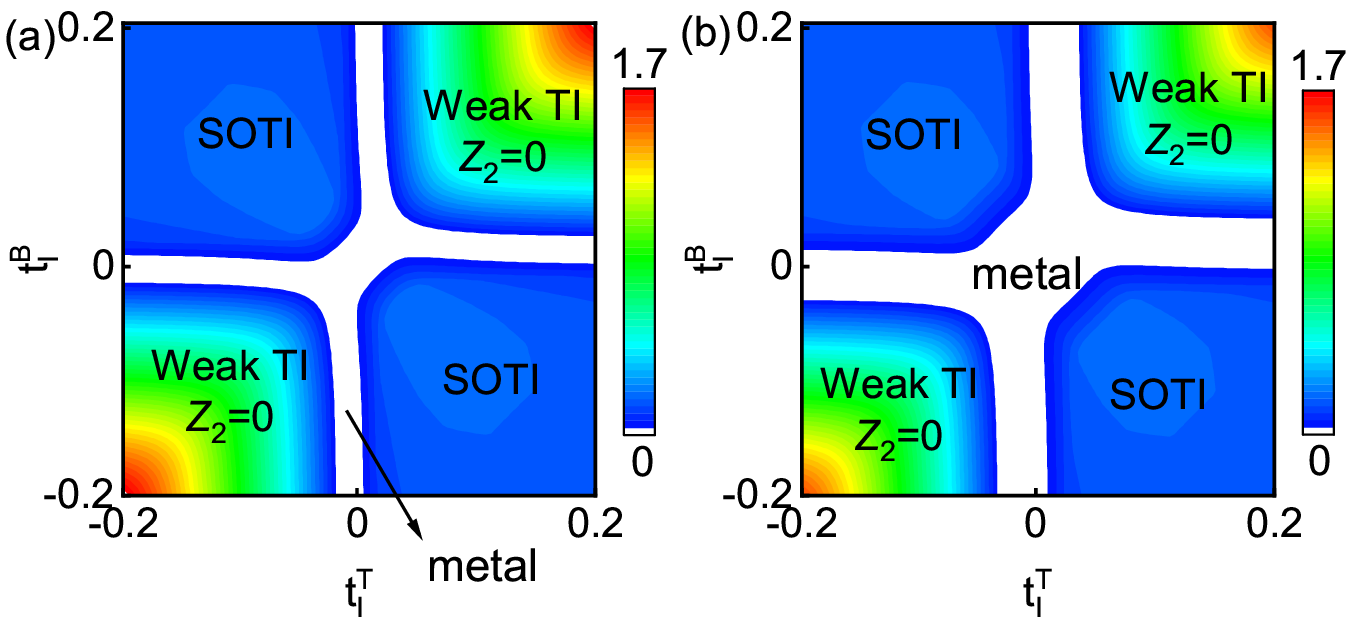}
  \caption{ Topological phase diagram of the coupled Kane-Mele models as functions of $t_{\rm I}^T$ and $t_{\rm I}^B$ at different Rashba spin-orbit coupling for (a) $t_{\rm R}=0.05$ and for (b) $t_{\rm R}=0.1$. Other parameters are set to be $t=1, \lambda_{T,B}=0, \eta=0.1$. Color measures the bulk gap amplitude in the Weak TI phases and the edge gap amplitude in the second-order TI phases.
  White areas correspond to metal phases.
  }
  \label{figpra}
\end{figure}

In Sec.~\ref{SectionII}, we have discussed that the corner states in the coupled Kane-Mele are robust against Rashba spin-orbit couplings.
To further demonstrate the effect of Rashba spin-orbit coupling on the phase diagram, we show the topological phase diagrams for the coupled Kane-Mele models within the parameter spaces of $t_{\rm I}^T$ and $t_{\rm I}^B$ at various Rashba spin-orbit coupling in Fig.~\ref{figpra}.
In Figs.~\ref{figpra}(a) and~\ref{figpra}(b) we set the $t_{\rm R}=0.05$ and $t_{\rm R}=0.1$, respectively, and the other parameters are set to $t=1, \lambda_{T,B}=0, \eta=0.1$. Comparing with Fig.~\ref{fig5}(a), we find that because the Rashba spin-orbit coupling leads to spin cleavage, the phase boundary broadens, leading to the formation of a metallic phase. As the Rashba spin-orbit coupling increases, the metallic phase region becomes larger. Notably, this does not destroy the higher-order topological insulator phase. These together suggest that the nontrivial higher-order topological insulator phase can also survive in the buckled honeycomb lattice.

\section{Generalization of the second-order TI formation mechanism to Bernevig-Hughes-Zhang models}\label{SectionIV}
In the above discussions, the main findings are all obtained in graphene-based topological systems. To show its universal characteristics, we examine the possibility of engineering second-order TI via interlayer coupling in the seminal Bernevig-Hughes-Zhang models. The Hamiltonian of the coupled Bernevig-Hughes-Zhang model systems can be written as:
\begin{align}
 H(\textbf{k})=\left( \begin{matrix}
 H_T(\textbf{k}) + B_z \sigma_z s_z &  \eta \\
 \eta ^*  &  H_B(\textbf{k})+ B_z \sigma_z s_z\\
\end{matrix} \right),
 \label{eq10}
\end{align}
where $H_T(\textbf{k})$ and $H_B(\textbf{k})$ are respectively the Hamiltonians of the top and bottom Bernevig-Hughes-Zhang models~\cite{Bernevig2006} with different settings. $\eta$ measures the interlayer coupling. $B_z$ is the out-of-plane Zeeman field that breaks the time-reversal symmetry and drives the Bernevig-Hughes-Zhang model to be a quantum anomalous Hall phase~\cite{Liu2008}. In momentum space, the Hamiltonian of a single Bernevig-Hughes-Zhang model system can then be written as $H_0=\sum_k \Psi_k^{\dag} H_0(k) \Psi_k$ on the basis of $\Psi_k = (\psi_{k11}, \psi_{k1\bar{1}}, \psi_{k\bar{1}1}, \psi_{k\bar{1}\bar{1}} )$, where $\psi_{k \sigma s} (\psi_{k \sigma s}^{\dag})$ destroys (creates) an electron with in-plane momentum $\textbf{k}=(k_x, k_y)$, orbital degree of freedom $\sigma \in \left \{ 1, \bar{1} \right \}$, and spin $s \in \left \{ 1, \bar{1} \right \}$. Then, $H_T(\textbf{k})$ and $H_B(\textbf{k})$ can be written as:
\begin{align}
 H_T(\textbf{k})=(t k_x^2 + t k_y^2 + \epsilon)\sigma_z + \lambda_x k_x \sigma_x s_z + \lambda_y k_y \sigma_y,\nonumber  \\
  H_B(\textbf{k})=(t k_y^2 + t k_x^2 + \epsilon)\sigma_z + \lambda_x k_y \sigma_x s_z + \lambda_y k_x \sigma_y,
 \label{eq11}
\end{align}
where $\sigma_i$ and $s_i$ for $i \in \left \{x, y, z\right\}$ are Pauli matrices acting on orbital and spin degrees of freedom, respectively.
$t$ is the nearest-neighbor intra-orbital hopping, $\epsilon$ describes a relative energy shift between the particle-like $(\sigma = 1)$ and hole-like $(\sigma = \bar{1})$ bands, and $\lambda_{x,y}$ measures the kinetic energy.

\subsection{Coupling the Bernevig-Hughes-Zhang model based $\mathbb{Z}_2$ TI}
\begin{figure}
  \centering
  \includegraphics[width=8.8cm,angle=0]{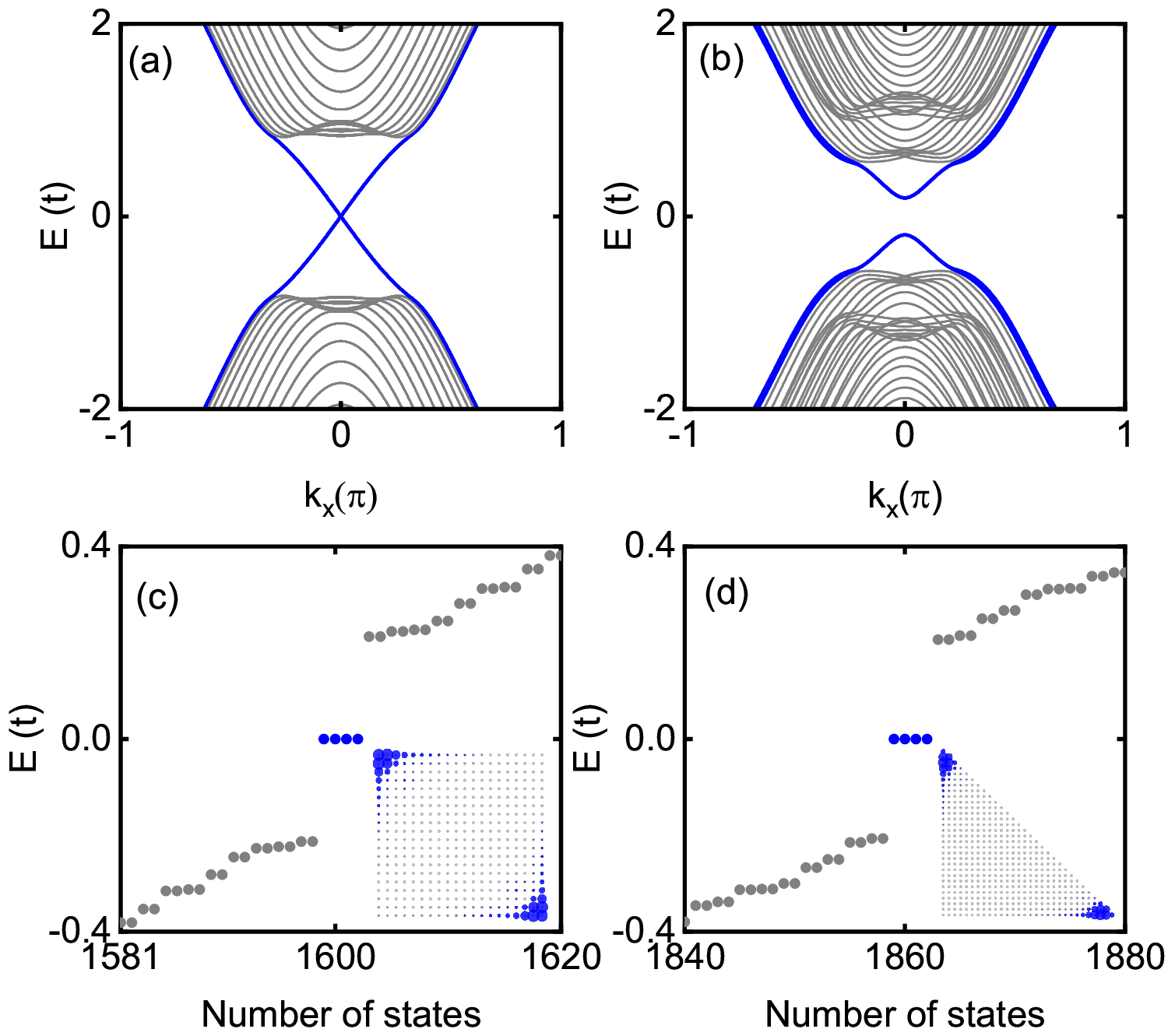}
  \caption{(a)-(b) Band structures of the nanoribbons for the coupled Bernevig-Hughes-Zhang model based $\mathbb{Z}_2$ TIs with (a) $\eta = 0$ and (b) $\eta = 0.3$. (c) Energy levels of square-shaped coupled nanoflake. (d) Energy levels of right triangle-shaped coupled nanoflake. Inset: The probability distribution of the corner state. The blue lines in (a) and (b) correspond to the spin-helical edge states. Other parameters are respectively $t = 1, \epsilon = -1, \lambda_x = \lambda_y = 1, B_z=0$, and the nanoribbon width $N_y = 20a$ for (a) and (b). The nanoflake size $20a \times 20a$ for (c), and $30a \times 30a$ for (d).}
  \label{fig6}
\end{figure}

Be setting $B_z = 0$, each layer is a representative $\mathbb{Z}_2$ TI. At the decoupled case, i.e., $\eta = 0.0$, Fig.~\ref{fig6}(a) displays the band structure of the decoupled $\mathbb{Z}_2$ TI. In the top and bottom layers, the same spin-polarized edge states (depicted by blue lines) propagate along opposite directions. Here, the parameters are set to be $t = 1, \epsilon = -1, \lambda_x = \lambda_y = 1$, and the nanoribbon width $N_y = 20a$, with $a$ representing the lattice constant. When the interlayer coupling is introduced, e.g., $\eta = 0.3$, the gapless spin-helical edge states become gapped (indicated by blue lines), as illustrated in Fig.~\ref{fig6}(b).

To verify whether the opened edge-state gap is topologically trivial or nontrivial, we calculate the energy levels of the rectangular samples with dimensions $20a \times 20a$. In Fig.~\ref{fig6}(c), four zero-energy in-gap states emerge, highlighted by the blue dots. The probability distributions of the wave functions for the in-gap states at half-filling are displayed in the inset of Fig.~\ref{fig6}(c). One can see that the half-electron charge of each state is localized at each corner, leading to the fractionalized charge distribution. This observation also strongly indicates that the interlayer coupling induces a second-order TI.

\begin{figure}
  \centering
  \includegraphics[width=8.8cm,angle=0]{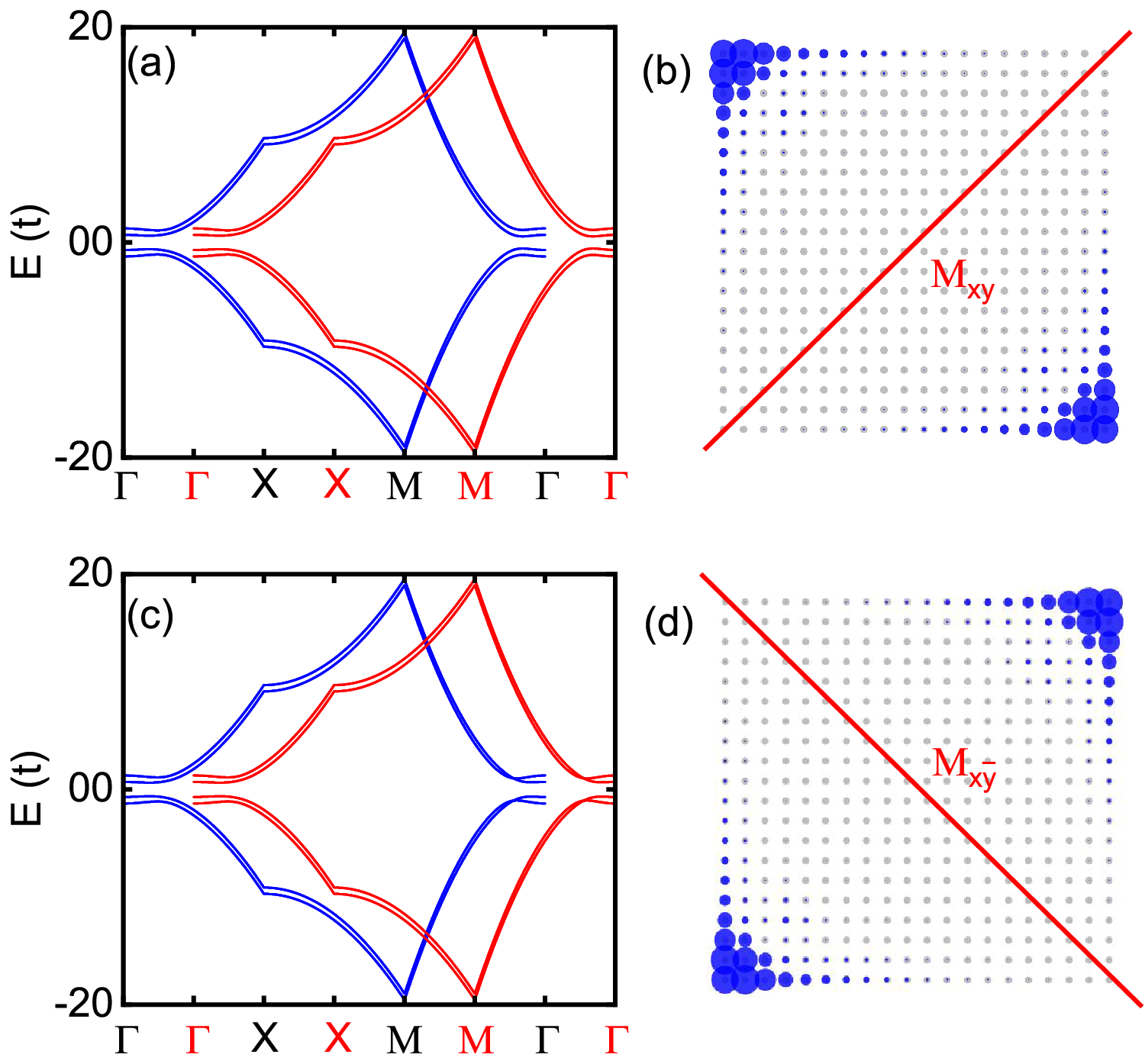}
  \caption{(a) Bulk band structures for $H(k_x, k_y)$ (blue line) and $H(k_y, k_x)$ (red line) by Eq.~(\ref{eq10}) along the high-symmetry line $\Gamma- X - M - \Gamma$ of the Brillouin zone for the coupled Bernevig-Hughes-Zhang $\mathbb{Z}_2$ TIs. (b) Probability distribution of corner states is protected by the mirror symmetry $\mathcal{M}_{xy}$. (c) Bulk band structures for $H(k_x, k_y)$ (blue line) and $H(-k_y, -k_x)$ (red line) by Eq.~(\ref{eq13}) along the high-symmetry line $\Gamma- X - M - \Gamma$ of the Brillouin zone for the coupled Bernevig-Hughes-Zhang $\mathbb{Z}_2$ TIs. (d) Probability distribution of corner states is protected by mirror symmetry $\mathcal{M}_{x\overline{y}}$. The parameters are the same as those in Fig.~\ref{fig6}(b).
}
  \label{fig7}
\end{figure}

To elucidate the origin of corner states, we analyze the symmetry properties of the system. First, we demonstrate that the Hamiltonian $H(\textbf{k})$ maintains time-reversal symmetry $\mathcal{T} = i s_y \mathcal{K}$, where $\mathcal{K}$ denotes the complex conjugation, i.e., $\mathcal{T} H(\textbf{k}) \mathcal{T}^{-1} = H(-\textbf{k})$. Second, we introduce the real-space mirror symmetry operator $\mathcal{M}_{xy}$, which transforms $H(k_x, k_y)$ into $H(k_y, k_x)$. To verify that the system satisfies the mirror symmetry $\mathcal{M}_{xy}$, we calculate the bulk bands of $H (k_x, k_y)$ and $H(k_y, k_x)$ along the high-symmetry line $\Gamma - X - M - \Gamma$ of the Brillouin zone, denoted in blue and red, respectively [see Fig.~\ref{fig7}(a)]. It shows that the bulk bands remain invariant under the operation of $\mathcal{M}_{xy}$. Our analysis shows that the second-order TI phase is protected by the mirror symmetry $\mathcal{M}_{xy}$ and the time-reversal symmetry, as shown in Fig.~\ref{fig7}(b). To emphasize the role of mirror symmetry, we show the energy level and corner state wave function distributions for the right-angled triangular nanoflake in Fig.~\ref{fig6}(d). One can see the corner state distribution in the two corners protected by the mirror symmetry $\mathcal{M}_{xy}$.

Based on the mirror-symmetry analysis, we again introduce the symmetry operator $\mathcal{M}_{x\overline{y}}$, which turns $H (k_x, k_y)$ into $H (-k_y, -k_x)$. We alter the stacking arrangement of the coupled system, i.e., modifying the Hamiltonian $H_B(\textbf{k})$ as described in Eq.~(\ref{eq10}):
\begin{align}
  H_B^{II}(\textbf{k})=(t k_y^2 + t k_x^2 + \epsilon)\sigma_z - \lambda_x k_y \sigma_x s_z - \lambda_y k_x \sigma_y.
 \label{eq12}
\end{align}
Then, the new coupled Bernevig-Hughes-Zhang $\mathbb{Z}_2$ TI system can be described by the following Hamiltonian:
\begin{align}
 H^{II}(\textbf{k})=\left( \begin{matrix}
 H_T(\textbf{k})+ B_z \sigma_z s_z&  \eta \\
 \eta ^*  &  H_B^{II}(\textbf{k})+ B_z \sigma_z s_z\\
\end{matrix} \right).
 \label{eq13}
\end{align}

Figure.~\ref{fig7}(c) shows that the bulk band structure of $H^{\rm II} (k_x, k_y)$ (blue lines) along the high-symmetry line $\Gamma - X - M - \Gamma$ is identical to that of $H^{\rm II} (-k_y, -k_x)$ (red lines). This indicates that the system remains invariant under the mirror symmetry operator $\mathcal{M}_{x\overline{y}}$. Consequently, the two Kramers pairs of corner states are distributed in the lower-left and upper-right corners due to the mirror symmetry $\mathcal{M}_{x\overline{y}}$ [see Fig.~\ref{fig7}(d)]. Our results demonstrate that the Kramers pairs of corner states are protected by both time-reversal symmetry and the real-space mirror symmetry $\mathcal{M}_{xy}$ or $\mathcal{M}_{x\overline{y}}$. And, the positions of the corner states can be controlled by the stacking mode of the coupled system.
By setting up the Bernevig-Hughes-Zhang model with zigzag boundaries and ensuring $\mathcal{M}_x$ or $\mathcal{M}_y$ mirror symmetry, we can also obtain the second-order TI phase. This is the same effective Hamiltonian as that of the in-plane magnetic field-induced Bernevig-Hughes-Zhang model to second-order TI~\cite{Ren2020}. The second-order TI of the coupled Bernevig-Hughes-Zhang model is different from that of the coupled Kane-Mele model because the graphene zigzag boundary depends on the $A/B$ sublattice.

\subsection{Coupling the Bernevig-Hughes-Zhang model based quantum anomalous Hall effects}
\begin{figure}
  \centering
  \includegraphics[width=8.8cm,angle=0]{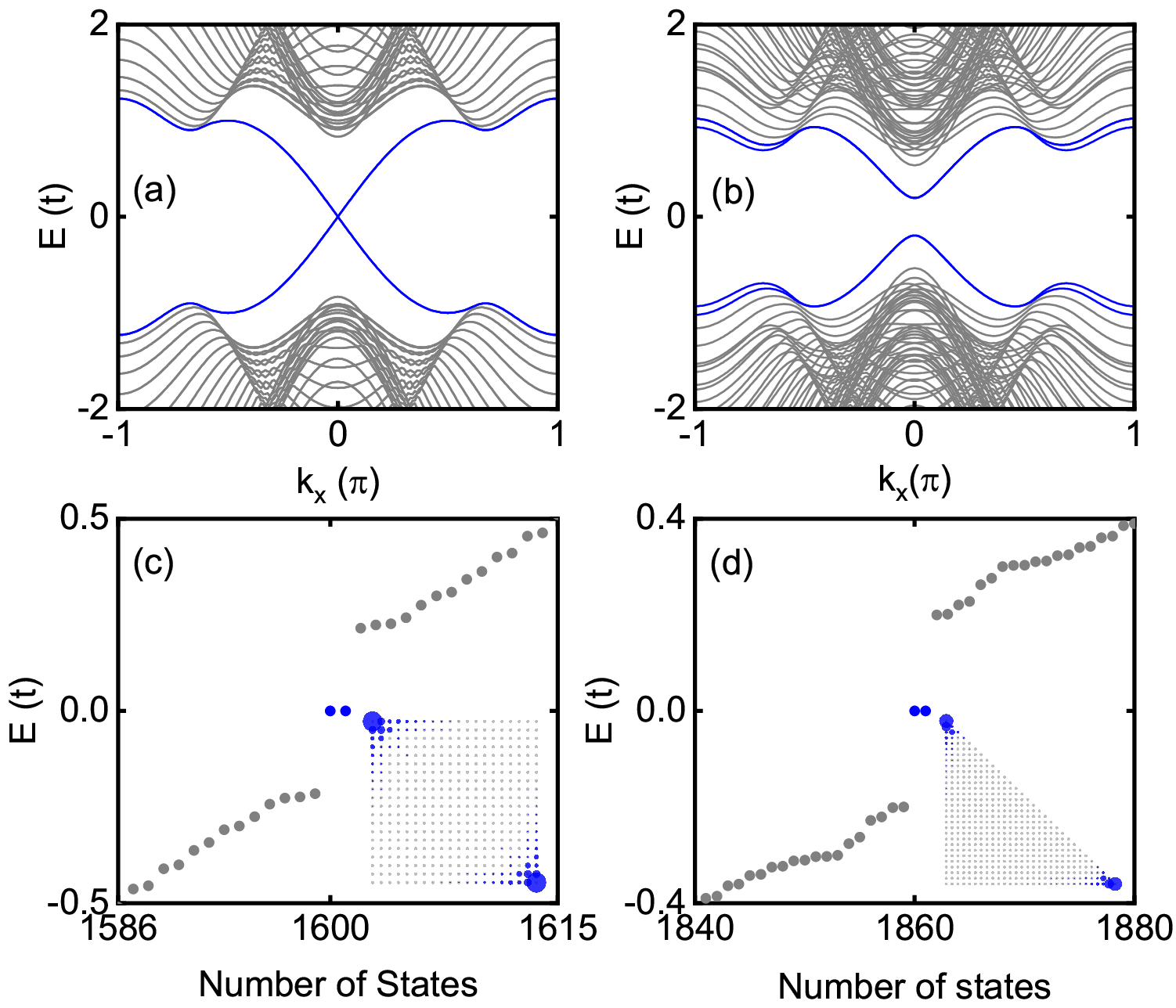}
  \caption{ (a)-(b) Band structures for the coupled Bernevig-Hughes-Zhang model based quantum anomalous Hall effects with (a) $\eta = 0$ and (b) $\eta = 0.3$. (c) Energy levels of square-shaped coupled nanoflake. (d) Energy levels of right triangle-shaped coupled nanoflake. Inset: The probability distribution of the corner state. The blue lines in (a) and (b)correspond to the chiral edge states. The system parameters are set to be $t = 1, \epsilon = -1, \lambda_x = \lambda_y = 1, B_z=0$, and the nanoribbon width $N_y = 20a$ for (a) and (b), the nanoflake size $20a \times 20a$ for (c), and $30a \times 30a$ for (d).}
  \label{fig8}
\end{figure}

We then extend our findings to another kind of quantum anomalous Hall effect, which can be obtained by introducing an out-of-plane Zeeman field in the Bernevig-Hughes-Zhang model~\cite{Liu2008}. The tight-binding Hamiltonian is expressed by Eq.~(\ref{eq10}), with an out-of-plane Zeeman field $B_z = 1.8$. To demonstrate the impact of interlayer coupling, we calculate the band structures of nanoribbons for the coupled system without and with interlayer coupling, as plotted in Figs.~\ref{fig8}(a) and~\ref{fig8}(b), respectively. In the absence of interlayer coupling, one can see two pairs of gapless chiral edge states respectively counterpropagate at the top and bottom layers in Fig.~\ref{fig8}(a). When a finite interlayer coupling is considered, the edge states become gapped [see Fig.~\ref{fig8}(a)], which signals the possible presence of a second-order TI.

In Fig.~\ref{fig8}(c), we display the energy levels of a square nanoflake with a boundary length of $L_n = 20a$. Other parameters are the same as those in Fig.~\ref{fig8}(b). One can see that only two zero-energy states are present (blue dots), which are localized at two corners of the square-shaped nanoflake, due to the mirror symmetry $\mathcal{M}_{xy}$, as shown in the inset of Fig.~\ref{fig8}(c). Furthermore, we also consider a right triangle nanoflake with two corner states existing at the two mirror symmetry corners, as shown in Fig.~\ref{fig8}(d). Thus, one can say that the interlayer coupling is also able to engineer the hybrid structure of two Bernevig-Hughes-Zhang model-based quantum anomalous Hall systems to be a second-order TI. Using the same parameters, we apply Eq.~(\ref{eq13}), which is protected by the mirror symmetry $\mathcal{M}_{x\overline{y}}$, to redistribute the corner states to the remaining two corners. Unlike the coupled Bernevig-Hughes-Zhang model based on $\mathbb{Z}_2$ TIs in Fig.~\ref{fig6}, the out-of-plane Zeeman field breaks the time-reversal symmetry. However, there is no breaking of the real-space mirror symmetry $\mathcal{M}_{xy}$ and $\mathcal{M}_{x\overline{y}}$. As a consequence, the interlayer coupling induces an edge state gap, and the mirror symmetry $\mathcal{M}_{xy}$ or $\mathcal{M}_{x\overline{y}}$ ensures the existence of corner states, i.e., the two-dimensional second-order TI.

\section{three-dimensional second-order topological semimetals}\label{SectionV}

\begin{figure} 
  \centering
  \includegraphics[width=8.6cm,angle=0]{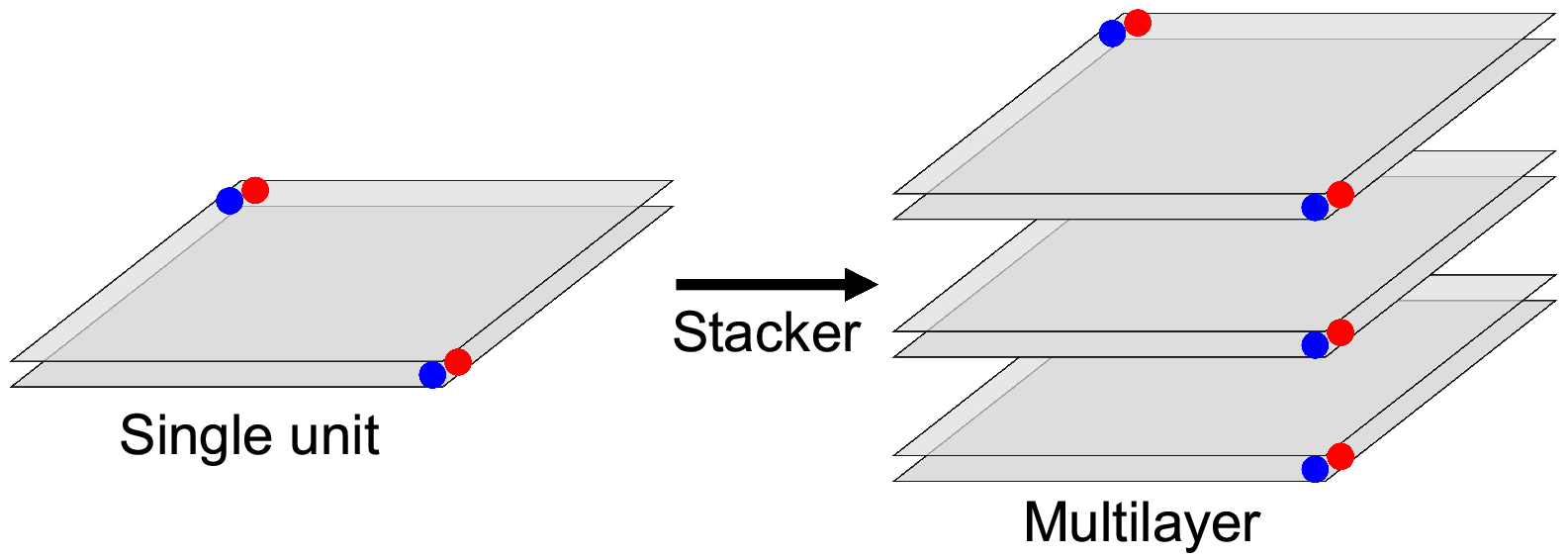}
  \caption{Schematic plot of stacked second-order TIs. Left: A unit of second-order TI with time-reversal symmetry, two Kramer pairs of corner states denoted by red and blue dots. Right: Stacked coupled systems. Only three superlattices are displayed.
  }
  \label{fig9}
\end{figure}

To address the second question raised in Section I, we explore the $k_z$-direction stacked time-reversal invariant second-order TIs (coupled Kane-Mele or Bernevig-Hughes-Zhang models) as shown in the left panel of Fig.~\ref{fig9}. This representation illustrates the coupled system as a supercell. In the right panel of Fig.~\ref{fig9}, we present the three-dimensional system obtained by stacking along the $k_z$-direction, with a stacking coupling strength of $t_\perp$, displaying only three supercells. Our analysis reveals that the second-order nodal ring semimetals can be realized in both the coupled Bernevig-Hughes-Zhang model and the Kane-Mele model after stacking. In addition, the three-dimensional Bernevig-Hughes-Zhang model exhibits a richer array of second-order topological phases, such as second-order Dirac semimetals.

\subsection{three-dimensional second-order nodal ring semimetals by using coupled Kane-Mele models}
\begin{figure}
  \centering
  \includegraphics[width=8.6cm,angle=0]{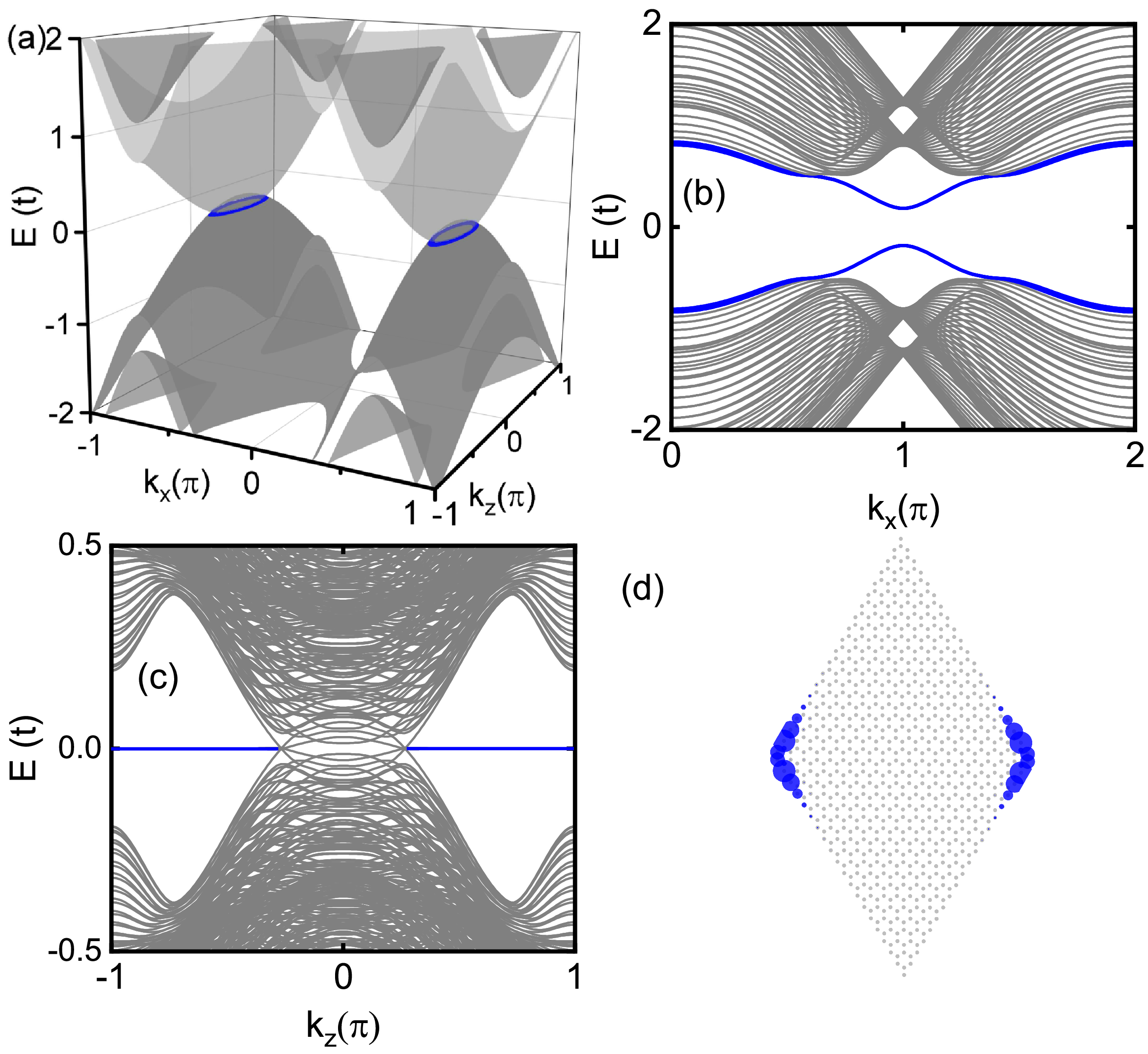}
  \caption{Electronic structure of the second-order nodal ring semimetal described by stacked coupled Kane-Mele models. (a) The bulk energy spectrum in the $k_x-k_z$ plane at $k_y=0$. The blue lines indicate a Dirac-type nodal ring. (b) The energy spectrum as a function of $k_x$ with the open boundary condition along the $y$ direction at $k_z = \pi$. The blue lines correspond to the gapped helical edge states.
  (c) The energy spectrum as a function of $k_z$ for a diamond-shaped sample with open boundary conditions along both $x$ and $y$ directions. The blue solid lines represent the four-fold degeneracy hinge Fermi arc states. (d) The probability distribution of the hinge Fermi arc states at $k_z = \pi$. The parameters are set to be $t = 1, t_{\rm I}^T = -t_{\rm I}^B = 0.1, t_{\rm R}=0, \lambda_{T,B}=0, \eta = 0.65, t_z = 0.45$.
  }
  \label{fig10}
\end{figure}

We first stack the coupled Kane-Mele model along the $k_z$-direction. The corresponding tight-binding Hamiltonian is given by:
\begin{align}
 H_{3DKM}=\left( \begin{matrix}
 H_T &  \eta + t_\perp e^{i k_z}\\
 \eta + t_\perp e^{-i k_z}  &   H_B\\
\end{matrix} \right),
 \label{eq14}
\end{align}
where $\eta$ represents the interlayer coupling between Kane-Mele models, and $t_z$ denotes the stacking coupling between superlattices (i.e., coupled Kane-Mele models) along the $k_z$ direction. $H_T$ and $H_B$ are represented by Eq.~(\ref{eq5}). We set the parameters as $t = 1, t_{\rm I}^T = - t_{\rm I}^B = 0.1, t_{\rm R}=0, \lambda_T= \lambda_B = 0$, and $\eta = 0.65$. When $t_\perp = 0$, Eq.~(\ref{eq14}) describes the time-reversal invariant two-dimensional second-order TI discussed in Section II. We stack the second-order TI with a coupling strength of $t_\perp = 0.45$ along the $k_z$-direction to obtain a three-dimensional system, as displayed in Fig.~\ref{fig9}. To elucidate the electronic properties of the system, we plot the bulk band structure as a function of $k_x-k_z$ at $k_y=0$ in Fig.~\ref{fig10}(a). Here, one can observe the bulk nodal ring (depicted by blue lines) lying in the mirror plane $k_z=0$ of the Brillouin zone. This observation confirms the presence of a nodal ring semimetal phase.

To illustrate the topological properties of the three-dimensional nodal ring semimetal, we calculate the energy band structure of the diamond-shaped graphene-flake with zigzag boundaries at $k_z = \pi$, as depicted in Fig.~\ref{fig10}(b). One can notice that the spin-helical edge states become gapped, as indicated by the blue lines. This observation aligns with previous findings for diamond-shaped coupled Kane-Mele nanoflakes, the gap-opening of the edge states serves as an indicator of the formation of second-order TI phases. Therefore, the two-dimensional subsystem in the three-dimensional system exhibits a second-order TI phase at fixed $k_z$. Furthermore, in Fig.~\ref{fig10}(c), we plot the energy spectrum along the $k_z$ direction for the diamond-shaped sample with open zigzag boundaries. One can notice the presence of hinge Fermi arc states characterized by zero-energy flat bands (blue line) that terminate on the projection of the nodal ring. Considering the time-reversal symmetry $\mathcal{T}$ and the effective mass analysis of edge states, it is evident that the fourfold degeneracy Fermi arc hinge states are located at the two obtuse corner hinges, as depicted in Fig.~\ref{fig10}(d). This suggests the formation of the second-order nodal ring semimetals.
One can conclude that the second-order nodal ring semimetals can be engineered by stacking along the $k_z$-direction using the coupled Kane-Mele models as a supercell.

\subsection{Three-dimensional second-order semimetals by using the coupled Bernevig-Hughes-Zhang models}
In this section, as a generalization of the second-order semimetals, we stack the coupled Bernevig-Hughes-Zhang models along the $k_z$-direction.
The corresponding tight-binding Hamiltonian can be expressed as:
\begin{align}
 H_{3D}(\textbf{k})=\left( \begin{matrix}
 H_T(\textbf{k}) &  \eta + t_\perp  e^{i k_z} \\
 \eta + t_\perp  e^{-i k_z}  &  H_B(\textbf{k}) \\
\end{matrix} \right),
 \label{eq15}
\end{align}
where $\eta$ represents the interlayer coupling of the coupled Bernevig-Hughes-Zhang model, and $t_\perp$ denotes the stacking amplitude along the $k_z$ direction. $H_T(\textbf{k})$ and $H_B(\textbf{k})$ are represented by Eq.~(\ref{eq11}). Significantly, the system is still protected by time-reversal symmetry, i.e., $\mathcal{T} H_{3D}(\textbf{k}) \mathcal{T}^{-1} = H_{3D}(-\textbf{k})$.

\subsubsection{Second-order nodal ring semimetal}
\begin{figure}
  \centering
  \includegraphics[width=8.6cm,angle=0]{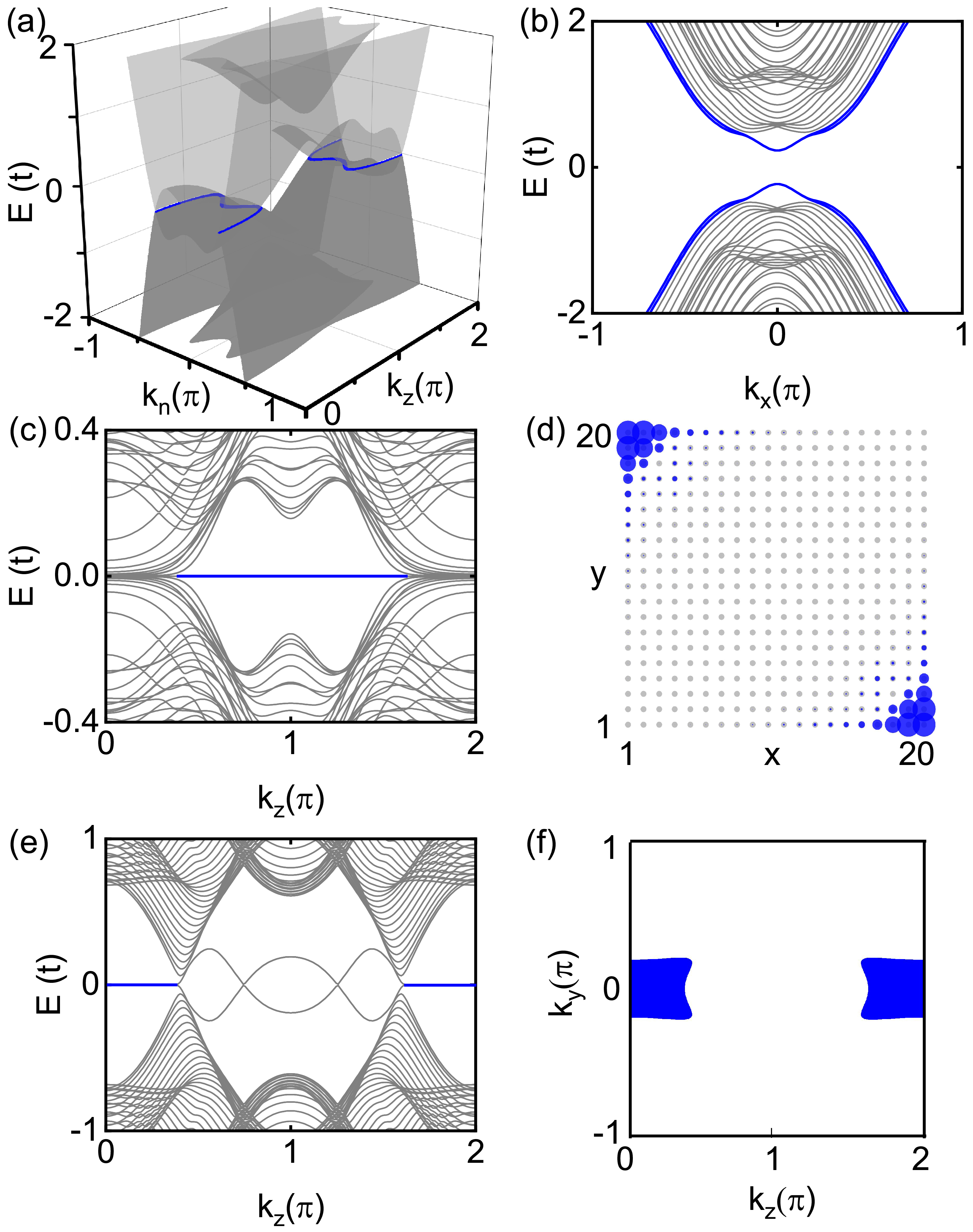}
  \caption{Electronic structure of the second-order nodal ring semimetal. (a) The bulk energy spectrum in the $k_n-k_z$ plane with $k_n$ along the $k_x = k_y$ direction. The blue lines indicate a Dirac-type nodal ring. (b) The energy spectrum as a function of $k_x$ with the open boundary condition along the $y$ direction for $k_z = 1.2\pi$. The blue lines correspond to the gapped helical edge states. (c) The energy spectrum as a function of $k_z$ with the open boundary conditions along both x and y directions. The blue solid lines represent the fourfold degeneracy hinge Fermi arc states. (d) The probability distribution of the hinge Fermi arc states of $k_z = 1.2\pi$. (e) The energy spectrum as a function of $k_z$ with the open boundary condition along the x direction for $k_y = 0$. The blue lines represent the drumhead surface states. (f) The spectral density, in the surface Brillouin zone defined in the $k_y-k_z$ plane at $E = 0$. The drumhead surface states appear in the regions bounded by the projected nodal ring. The parameters are chosen as $t = 1, \epsilon=-1 , \lambda_x = \lambda_y = 1, \eta = 0.7, t_z = 0.5$.
  }
  \label{fig11}
\end{figure}

First, we consider the case where the parameters of the interleaved layer TIs are the same. We set the parameters $\epsilon =-1, t = 1, \lambda_x = \lambda_y = 1$, and $\eta = 0.7$. When $t_\perp = 0$, Eq.~(\ref{eq15}) describes the time-reversal invariant two-dimensional second-order TI in the left panel of Fig.~\ref{fig9}.
We stack the second-order TI with a coupling strength of $t_\perp = 0.5$ along the $k_z$-direction to obtain a three-dimensional system, as shown in the right panel of Fig.~\ref{fig9}.
To illustrate the electronic properties of the system, in Fig.~\ref{fig11}(a) we plot the bulk band structure in the planes of $k_n-k_z$, where $k_n = k_x = k_y$. One can see the bulk nodal ring (depicted by blue lines) lying in the mirror plane $k_x=k_y$ of the Brillouin zone and other band structures around them. This observation confirms that the system is a nodal ring semimetal.

\begin{figure}[htbp]
  \centering
  \includegraphics[width=8.8cm,angle=0]{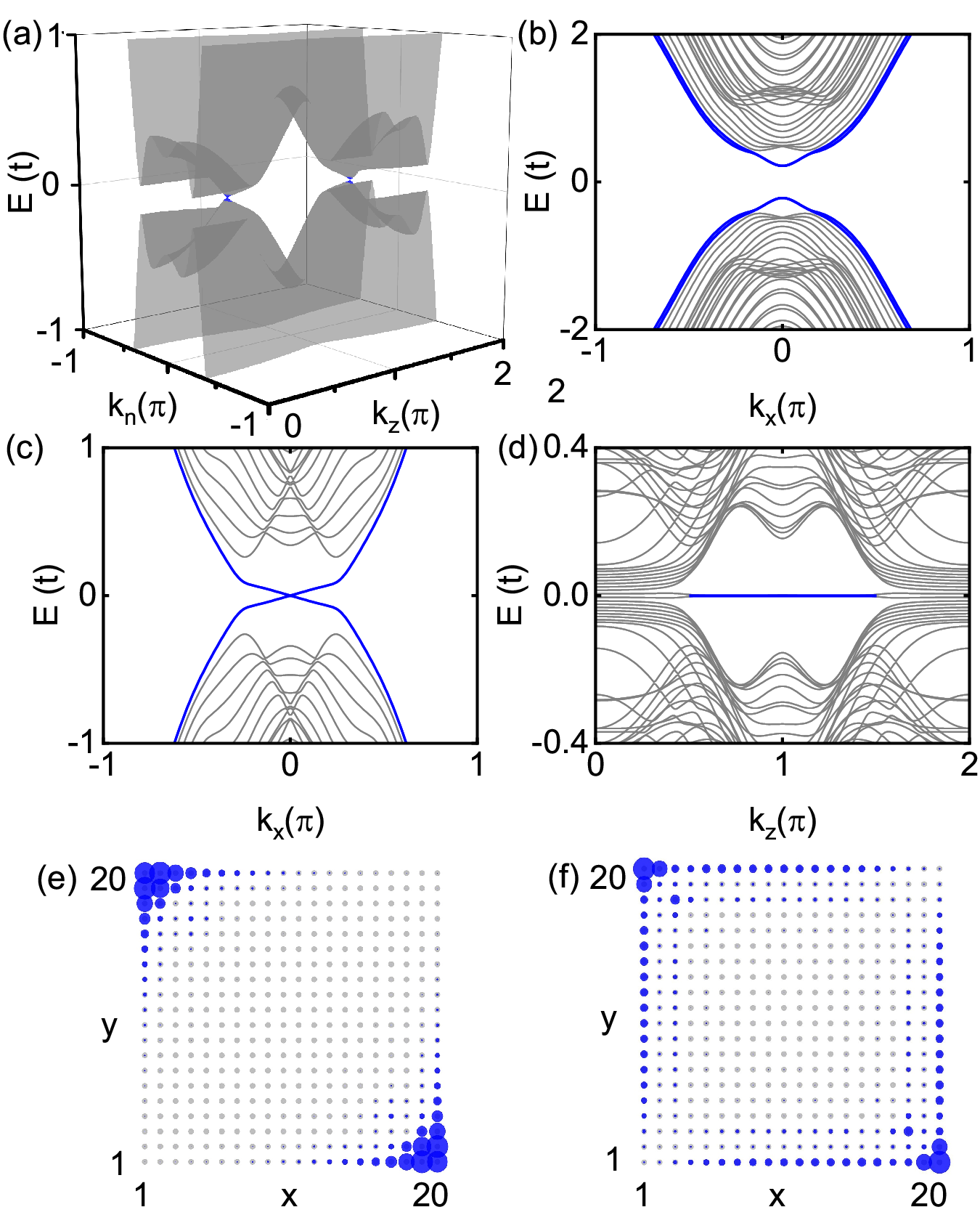}
  \caption{Electronic structure of the second-order Dirac semimetal. (a) The bulk energy spectrum in the $k_n-k_z$ plane with $k_n$ along the $k_x = k_y$ direction. The blue points indicate Dirac nodes. (b) The energy spectrum as a function of $k_x$ with the open boundary condition along the $y$ direction at $k_z = 1.2\pi$. (c) The energy spectrum as a function of $k_x$ with the open boundary condition along the $y$ direction at $k_z = 0$. The blue lines in (b) and (c) correspond to the edge states. (d) The energy spectrum as a function of $k_z$ with the open boundary conditions along both $x$ and $y$ directions. The blue solid lines represent the fourfold degeneracy hinge Fermi arc states. (e) The probability distribution of the hinge Fermi arc states at $k_z = 1.2\pi$. (f) The probability distribution of the surface Fermi arc states at $k_z = 0$. Here, we set $\epsilon_T =-1, \epsilon_B =-0.8$. Other parameters are the same as in Fig.~\ref{fig11}.
  }
  \label{fig12}
\end{figure}
To demonstrate the topological features of the three-dimensional nodal ring semimetal, we set the open boundary along the $y$ direction.
The energy spectrum as a function of $k_x$ is shown in Fig.~\ref{fig11}(b) at $k_z = 1.2\pi$, corresponding to the bulk gap region of Fig.~\ref{fig11}(a). The gapped helical edge state is highlighted with blue lines, serving as a signal for the formation of second-order TI. Figure~\ref{fig11}(c) plots the energy spectrum along the $k_z$ direction for the system with open boundaries in both $x$ and $y$ directions.
One can see that the dispersionless hinge Fermi arc states with zero-energy flat bands (blue lines) terminate on the projection of the nodal ring. Due to the time-reversal symmetry $\mathcal{T}$ and the mirror symmetry $\mathcal{M}_{xy}$, the fourfold degeneracy Fermi arc hinge states are located at the two mirror-symmetric diagonal hinges, as shown in Fig.~\ref{fig11}(d). This observation serves as a signature of the second-order nodal ring semimetals. Additionally, we confirm the presence of drumhead surface states in second-order nodal ring semimetals.
In Fig.~\ref{fig11}(e), by setting the $x$ direction as an open boundary and calculating the energy spectrum along the $k_z$ direction at $k_y = 0$, we observe eigenvalues of drumhead surface states (highlighted by blue lines) pinned to zero energy. Therefore, the second-order nodal ring semimetal also supports flat-band drumhead surface states within the region enclosed by the projections of the nodal ring onto the surface Brillouin Zone, as shown in Fig.~\ref{fig11}(f).

\subsubsection{Second-order Dirac semimetal}
Now, we introduce different relative orbital energy shifts in staggered layers, specifically $\epsilon_T = -1$ and $\epsilon_B = -0.8$, while maintaining other parameters being consistent with those in Fig.~\ref{fig11}. Figure~\ref{fig12}(a) shows the bulk band spectrum in the $k_n-k_z$ plane with $k_n$ along the $k_x =k_y$ direction. It is observed that the bulk Dirac-type nodal ring transforms into two protected by time-reversal symmetry Dirac nodes (blue points) positioned along the line of $k_x = k_y = 0$. We commence our analysis by examining the electronic properties of a two-dimensional subsystem at fixed $k_z$. Figures~\ref{fig12}(b) and~\ref{fig12}(c) depict the energy spectrum along the $k_x$ direction at $k_z=1.2\pi$ and $k_z=0$, respectively, with open boundaries in the $y$ direction. Notably, at $k_z = 1.2\pi$, a gapped helical edge state is evident, in contrast with the gapless helical edge states observed at $k_z = 0$.
Therefore, the Dirac nodes divide the $k_z$-dependent two-dimensional plane in the Brillouin zone into the second-order TI phase and $\mathbb{Z}_2$ TI phase.

To illustrate the overall topological properties of the three-dimensional system, we present in Fig.~\ref{fig12}(d) the energy spectrum along the $k_z$ direction for the system with open boundaries in both $x$ and $y$ directions. Two Kramer pairs of hinge Fermi arc states with zero-energy flat bands (depicted by blue lines) terminate on the projection of Dirac nodes, which are characteristic features of second-order topology in Dirac semimetals. As depicted in Fig.~\ref{fig12}(e), these hinge Fermi arc states serve as second-order topological signatures, localized within mirror-symmetric hinges. In Fig.~\ref{fig12}(f), we analyze the distribution of states around the Fermi surface at $k_z = 0$, where the first-order surface Fermi arc states of the three-dimensional system are distinctly visible.
Here, the hinge Fermi arc states correspond to corner states within the two-dimensional subsystem, while the surface Fermi arc states correspond to edge states. As a consequence, the system represents a theoretical realization of a three-dimensional second-order Dirac semimetal with both first-order and second-order topological states.

\section{summary}\label{SectionVI}

In summary, we present a general strategy for realizing second-order TI, i.e., coupling two TIs with opposite topological invariants.
The TI includes $\mathbb{Z}_2$ TIs (Kane-Mele model and Bernevig-Hughes-Zhang model), $\mathcal{C}=1$ (Bernevig-Hughes-Zhang model with exchange field), and $\mathcal{C}=2$ (Graphene with Rashba spin-orbit coupling and exchange fields) quantum anomalous Hall effect system. We use the effective Hamiltonian to illustrate that interlayer coupling leads to edge-state gaps in TIs. We calculate the energy spectra and energy levels of various coupled systems, showing that the interlayer coupling opens the edge state gap to induce the second-order TIs (i.e., the zero-energy corner states). The corner states are protected by mirror symmetry and can be explained by the effective mass of the gapped edge states. Furthermore, we stack time-reversal invariant second-order TI (coupled Kane-Mele model and Bernevig-Hughes-Zhang model) along the $ k_z$ direction to obtain three-dimensional second-order nodal ring semimetals.
This second-order nodal ring semimetal supports both drumhead surface states and one-dimensional gapless hinge Fermi arc states.
Interestingly, by changing the orbital energy transfer of the semimetal stacked by the Bernevig-Hughes-Zhang model, the nodal ring evolves into two Dirac nodes, resulting in the formation of a second-order Dirac semimetal. The Dirac semimetals support both one-dimensional gapless hinge states and two-dimensional surface states.

This work was financially supported by the National Natural Science Foundation of China (Grants No. 12074097, No. 11974327, and No. 12004369), Anhui Initiative in Quantum Information Technologies (Grant No. AHY170000), and Innovation Program for Quantum Science and Technology (Grant No. 2021ZD0302800). The supercomputing service of USTC is gratefully acknowledged.

$\ddag$ These authors contribute equally to this work.


\end{document}